\documentclass{amsart}

\usepackage{enumerate}
\usepackage{color}

\usepackage{amssymb}
\usepackage{latexsym}
\usepackage{array}
\usepackage{graphics}
\usepackage{epstopdf}
\usepackage{epsf,graphicx,subfigure}
\usepackage{multirow}
\usepackage{hyperref} % lineno
\usepackage{tikz}
\usetikzlibrary{arrows,positioning,shapes.geometric}

\usepackage[english]{babel} % English language/hyphenation
\usepackage{amsmath,amsfonts,amsthm,bm} % Math packages
\usepackage[linesnumbered,commentsnumbered,ruled]{algorithm2e}
\usepackage{tabularx}
\usepackage{booktabs}

\definecolor{dark_grey}{gray}{0.7}

\newcolumntype{v}[1]{>{\raggedright\hspace{0pt}}p{#1}}
\newcolumntype{V}[1]{>{\scriptsize\raggedright\hspace{0pt}}p{#1}}

\begin{document}

\title[Sleep dynamics by unsupervised learning]{Explore intrinsic geometry of sleep dynamics and predict sleep stage by unsupervised learning techniques}

\author[G.-R. Liu]{Gi-Ren Liu}
\address{Department of Mathematics, National Chen-Kung University, Tainan, Taiwan}
\email{girenliu@ncku.edu.tw}

\author[Y.-L. Lo]{Yu-Lun Lo}
\address{Department of Thoracic Medicine, Healthcare Center, Chang Gung Memorial Hospital, Chang Gung University, School of Medicine, New Taipei, Taiwan}
\email{loyulun@hotmail.com}

\author[Y.-C. Sheu]{Yuan-Chung Sheu}
\address{Department of Applied Mathematics, National Chiao Tung University, Hsinchu, Taiwan}
\email{sheu@math.nctu.edu.tw}

\author[H.-T. Wu]{Hau-Tieng Wu}
\address{Department of Mathematics and Department of Statistical Science, Duke University, Durham, NC, USA. Mathematics Division, National Center for Theoretical Sciences, Taipei, Taiwan}
\email{hauwu@math.duke.edu}

\maketitle

\begin{abstract}
We propose a novel unsupervised approach for sleep dynamics exploration and automatic annotation by combining modern harmonic analysis tools. Specifically, we apply diffusion-based algorithms, diffusion map (DM) and alternating diffusion (AD) algorithms, to reconstruct the intrinsic geometry of sleep dynamics by reorganizing the spectral information of an electroencephalogram (EEG) extracted from a nonlinear-type time frequency analysis tool, the synchrosqueezing transform (SST).
The visualization is achieved by the nonlinear dimension reduction properties of DM and AD.
%
%To backup the proposed algorithm, we summarize the theoretical foundation of SST, DM and AD.
%
Moreover, the reconstructed nonlinear geometric structure of the sleep dynamics allows us to achieve the automatic annotation purpose. The hidden Markov model is trained to predict the sleep stage.
The prediction performance is validated on a publicly available benchmark database, Physionet Sleep-EDF [extended] SC$^*$ and ST$^*$, with the leave-one-subject-out cross validation.
The overall accuracy and macro F1 achieve $82.57\%$ and $76\%$ in Sleep-EDF SC$^*$ and $77.01\%$ and $71.53\%$ in Sleep-EDF ST$^*$, which is compatible {with} the state-of-the-art results by supervised learning-based algorithms.
The results suggest the potential of the proposed algorithm for clinical applications.
\end{abstract}

\section{Introduction}
Sleep is a recurring physiological dynamical activity in mammals. Since 1968, the Rechtschaffen and Kales (R\&K) criteria \cite{Rechtschaffen_Kales:1968} is the gold standard when researchers study human sleep dynamics, and this criteria was further generalized by American Academy of Sleep Medicine (AASM) \cite{Iber2007}.
According to the AASM criteria, the sleep dynamics is quantified by finite discrete stages, and those stages can be divided into two broad categories, the rapid eye movement (REM) and the non-rapid eye movement (NREM), and the NREM stage is further divided into shallow sleep (stages N1 and N2) and deep sleep (stage N3).
Based on this quantification, up to now, we have accumulated plenty of knowledge about sleep dynamics \cite{Saper2013} and sleep dynamics is nowadays an active research field due to more unknowns \cite{Kanda2016}.
Despite those unknowns, it has been well known that a distortion of sleep dynamics could lead to catastrophic outcomes. For example, REM disturbance slows down the perceptual skill improvement \cite{Karni1994}, insufficient N2 sleep is associated with weaning failure \cite{RocheCampo2010}, deprivation of slow wave sleep is associated with Alzheimer's disease \cite{Kang2009}, etc. Moreover, several public disasters are caused by low sleep quality \cite{Leger_Bayon_Laaban_Philip:2012}.
%Clearly, there are a lot of clinical applications from knowing the sleep dynamics.

The polysomnography (PSG) is the gold standard of evaluating the sleep dynamics. The PSG usually records multiple channels from a subject, ranging from electroencephologram (EEG), electrooculogram (EOG), electrocardiogram (ECG), electromyogram (EMG), photoplethysmogram (PPG), several respiratory signals, etc. While each channel contains its own  information about sleep, to apply the AASM criteria \cite{Iber2007} to study sleep dynamics, sleep experts mainly count on EEG, EOG and EMG. To simplify the discussion, and better explore how much sleep dynamics is captured by EEG signals, we focus on exploring sleep dynamics via analyzing EEG signals.

An EEG is a complicated time series. It is non-stationary in nature. A common model to study a given EEG assumes that it is {composed of} diverse spectra \cite{campbell2009eeg}, each of which depicts a portion of the brain dynamics. Furthermore, we assume that the brain dynamics, as a dynamical system, is supported on a low dimensional manifold. Based on this model, a natural question to ask is if we are able to recover the low dimensional manifold for the sleep dynamics study. The first step toward answering this question is to quantify the spectral content in an EEG signal by the time-frequency analysis.
Synchrosqueezing transform (SST) \cite{Daubechies_Lu_Wu:2011,Wu:2011Thesis} is a nonlinear-type time-frequency analysis technique, which allows us to combine the phase information into the spectrogram (the squared magnitude of the short-time Fourier transform (STFT)). Note that the phase information is ignored in the spectrogram. A direct consequence is ``sharpening'' the spectra of the EEG signal {and prevents the ``energy leakage''} that is commonly encountered in the spectrogram due to the uncertainty principle \cite{Ricaud2014}. Specifically, the spectrogram is sharpened by taking the phase information of STFT into account to nonlinearly deform the spectrogram, and the blurring effect/energy leakage of the spectrogram caused by the uncertainty principle is alleviated. We call the outcome the {\em synchrosqueezed EEG spectral features}, {which contains} not only the spectrogram, but also the EEG phase information.
The synchrosqueezed EEG spectral feature is in general different from the intrinsic sleep dynamics. This difference comes from various resources {including, for example, the distortion caused by the transform and inevitable noise and artifact.}
Thus, an extra step is needed to {refine/reorganize the synchrosqueezed EEG spectral feature and define the final intrinsic features for the} sleep dynamics.
We suggest to conquer the distortion by the local Mahalanobis distance (MD) framework \cite{singer2008non,TalmonPNAS}. Via local MD, we reorganize the synchrosqueezed EEG spectral features by the diffusion-based machine learning algorithms, including diffusion maps (DM) \cite{coifman2006diffusion}, alternating diffusion (AD) \cite{lederman2015alternating}, and co-clustering \cite{dhillon2001co}, depending on how many EEG channels we have, for this purpose.
We call the {resulting} features the {\em intrinsic sleep dynamical features}. Based on the established theory, the intrinsic sleep dynamical features recover the intrinsic geometry of sleep dynamics, and hence provides a visualization tool to explore sleep dynamics.

A direct application of discovering the geometry of sleep dynamics is an automatic annotation system. Scoring the overnight sleep stage from the PSG outputs by sleep experts is time consuming \cite{rotenberg2010wait} and error-prone \cite{Norman2000} due to the huge signal loading. Due to its importance for the whole healthcare system, an accurate automatic sleep dynamics scoring system is critical in the current clinical environment.
{For this automatic annotation purpose, we consider the standard hidden Markov model (HMM) \cite{fraser2008hidden} to learn the sleep experts' knowledge.}  Based on the physiological knowledge, we first take the available phenotype information of a new-arriving subject to determine ``similar subjects'' from the existing database. Then, based on the intrinsic sleep dynamical features and the annotations of these similar subjects, a prediction model for the new subject is established for the new-arriving subject.
To evaluate the performance of the proposed algorithm for the prediction purpose, we consider the publicly available benchmark database, the PhysioNet Sleep-EDF database \cite{Goldberger_Amaral_Glass_Hausdorff_Ivanov_Mark_Mietus_Moody_Peng_Stanley:2000}. This database contains two subsets, the SC* and ST*. The SC* subset consists of $20$ healthy subjects without any sleep-related medication, and the ST$^{*}$ subset consists of $22$ subjects who had mild difficulty falling asleep.
The overall accuracy and macro F1 achieve $82.57\%$ and $76\%$ in Sleep-EDF SC$^*$ and $77.01\%$ and $71.53\%$ in Sleep-EDF ST$^*$, which is compatible {with} the state-of-the-art results by supervised learning-based algorithms.
The results suggest the potential of the proposed algorithm for clinical applications.

The rest of this paper is organized as follows. In Section \ref{sec:SST}, we summarize the theoretical background of SST and demonstrate how SST works in an EEG signal. In Section \ref{Section:MetricDesignEIG}, the local MD and the empirical intrinsic geometry framework are summarized.
In Section \ref{sec:DM}, the diffusion-based algorithm, DM, is discussed and its theoretical support is summarized. In Section \ref{Section:Multiview}, diffusion-based sensor fusion algorithms, include AD, co-clustering and multiview DM, as well as known theoretical background, are provided.
In Section \ref{sec:proposed algorithm}, the implementation details of the proposed algorithm for recovering the geometry of sleep dynamics, including feature extraction algorithm for the synchrosqueezed EEG spectrum and diffusion-based feature organization and sensor fusions. The HMM for the automatic sleep stage annotation is also summarized.
In Section \ref{sec:description_database}, we describe the publicly available benchmark database, the Sleep-EDF Database [Expanded] from PhysioNet, and the statistics for the performance evaluation purpose.
In Section \ref{sec:experiments},
the results of applying the proposed algorithm to the Sleep-EDF Database [Expanded] are shown. The paper is closed with the discussion and conclusion in Section \ref{Section:Conclusion}, with a comparison with existing relevant literature in automatic sleep stage annotation.

\section{Synchrosqueezing transform}\label{sec:SST}

In this subsection, we introduce the synchrosqueezed EEG spectrogram based on the STFT-based SST \cite{Wu:2011Thesis} algorithm. The basic idea underlying SST is utilizing the phase information in the STFT of the EEG signal to sharpen the spectrogram. There are two benefits. First, the phase information that is commonly ignored in the spectrogram is preserved. Second, the uncertainty principle intrinsic to the spectrogram is alleviated and the spectrogram is sharpened, which can prevent the ``energy leakage'' caused by the blurring effect inherited in the uncertainty principle associated with the STFT \cite{Ricaud2014}. These two benefits allow us a better quantification of the EEG dynamics.
The SST algorithm is composed of three steps --  extract the local dynamics of the EEG signal by the STFT, manipulate the phase information, and sharpen the spectrogram according to the extracted phase.
Below we summarize the SST based on STFT.

Take a continuously recorded signal $f$, for example, an EEG in this work. In practice, $f$ can be as general as a tempered distribution function. Take a Schwartz function $h$ to be the chosen window. The STFT of $f$ is then defined by
\begin{equation}\label{STFT_continuous}
V^{(h)}_f(t,\omega)=\int_{-\infty}^\infty f(s) h(s-t)e^{-i2\pi \omega (s-t)} ds,
\end{equation}
where $t\in\mathbb{R}$ is the time, and $\omega\in\mathbb{R}$ is the frequency. We call $|V^{(h)}_f|^2:\mathbb{R}\times\mathbb{R}\to \mathbb{R}^+\cup\{0\}$ the {\em spectrogram}.
As a complex-valued function, the local phase information of $f$ could be approximated by $\partial_tV^{(h)}_f(t,\omega)$.
Intuitively, the phase function records the number of oscillations, and this intuition leads us to calculate the following  {\em reassignment rule}:
\begin{equation}\label{alogithm:sst:ressigment}
\Omega^{(h)}_f(t,\omega):=\left\{\begin{array}{ll}
\displaystyle \Im\frac{\partial_tV^{(h)}_f(t,\omega)}{2\pi V^{(h)}_f(t,\omega)} & \mbox{when}\quad |V^{(h)}_f(t,\omega)|\neq 0;\\
-\infty&\mbox{when}\quad |V^{(h)}_f(t,\omega)|=0\,,
\end{array}\right.
\end{equation}
where $\Im$ means taking the imaginary part.
The spectrogram of $f$ is finally sharpened by re-allocating the coefficients of the spectrogram according to the reassignment rule \cite{Berrian_Saito:2017}:
\begin{equation}\label{alogithm:sst:formula}
S^{(h)}_f(t,\xi):=\int |V^{(h)}_f(t,\omega)|^2\frac{1}{\alpha}g\Big(\frac{|\omega-\Omega^{(h)}_f(t,\omega)-\xi|}{\alpha}\Big) d\omega\,,
\end{equation}
where $\xi>0$ is the frequency and $g$ is a Schwartz function so that $g(\cdot/\alpha)/\alpha$ converges weakly to the Dirac measure supported at $0$ when $\alpha\to 0$. Theoretically, $\alpha$ controls the resolution of the frequency axis in the SST. This seemingly complicated transform has an intuitive interpretation. At each time $t$, we identify all spectrogram entities that contain oscillatory components at frequency $\xi$ by reading $\frac{1}{\alpha}g\big(\frac{|\omega-\Omega^{(h)}_f(t,\omega)-\xi|}{\alpha}\big)$, and put all identified entities in the $(t,\xi)$ slot. We call $S^{(h)}_f(t,\xi)$ the {\em synchrosqueezed spectrogram}.

{Note that the SST defined in \ref{alogithm:sst:formula} is slightly different from that introduced in \cite{Wu:2011Thesis} -- in \cite{Wu:2011Thesis}, it is the STFT that is re-allocated, but here, it is spectrogram that is re-allocated. We choose to re-allocate the spectrogram since we do not need to reconstruct the components in this work, and it is the sharpened energy distribution that encodes the phase information that we are interested in.}
In addition to providing a sharp and concentrated spectrogram, it has been proved in \cite{Chen_Cheng_Wu:2014} that SST is robust to different kinds of noise. This property is desirable since the EEG signal is commonly noisy. For theoretical developments and more discussions, we refer readers to \cite{Daubechies_Lu_Wu:2011,Chen_Cheng_Wu:2014}.\\

\noindent
{\it Example.}
To have a better insight of the SST algorithm, look at the following {\em toy} example that motivates the design of SST. Take a harmonic function $f(t)=e^{i2\pi \omega_{0}t}$, where $\omega_{0}>0$ is constant. By a direct calculation, we have $V^{(h)}_f(\omega,t)=e^{i2\pi \omega_{0} t}\exp (-2\pi^{2} (\omega-\omega_{0})^{2} H^{2} )$. From (\ref{alogithm:sst:ressigment}), we know that the instantaneous frequency of $f(t)$ can be recovered from the phase information of $V^{(h)}_f(\omega,t)$; that is $\omega_{0}=\frac{1}{2\pi}\Im\big(\frac{\partial}{\partial t}\ln V^{(h)}_{f}(t,\omega)\big)$. A direct calculation leads to $\omega_{0}=\omega-\textup{Im}\big(\frac{1}{2\pi H}\frac{\mathbf{S}_{g'}(\omega,t)}{\mathbf{S}_{g}(\omega,t)}\big)$, which means that the spectra energy near $\omega$ is spread from $\omega-\textup{Im}\big(\frac{1}{2\pi H}\frac{\mathbf{S}_{g'}(\omega,t)}{\mathbf{S}_{g}(\omega,t)}\big)$ for any $t>0$. Thus, to obtain a sharp spectrogram, we only need to reallocate the spectrogram to the right frequency $\omega_0$.

\begin{figure}
\centering
\includegraphics[scale=0.47]{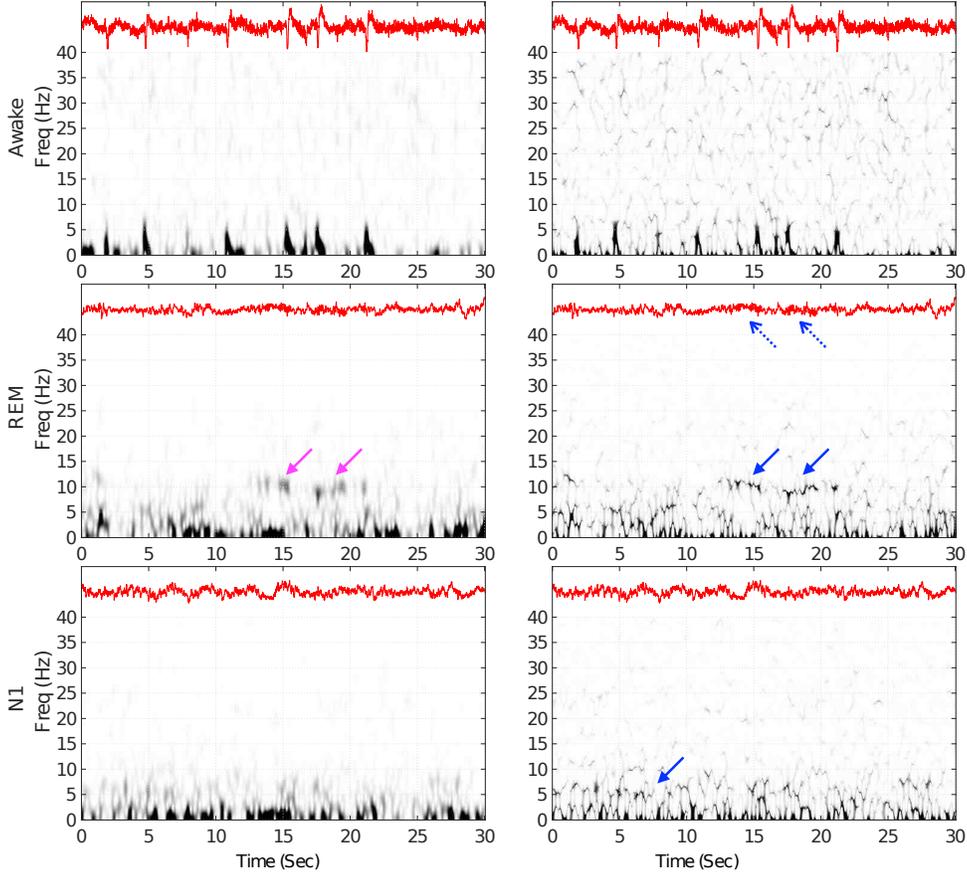}
\caption{An illustration of EEG spectrogram (left) and synchrosqueezed EEG spectrogram (right) of sleep stages Awake, REM and N1 from Sleep-EDF SC$^*$ database. The EEG signals are superimposed as red curves for visual comparison.
}
\label{Figure0}
\end{figure}

\begin{figure}
\centering
\includegraphics[scale=0.47]{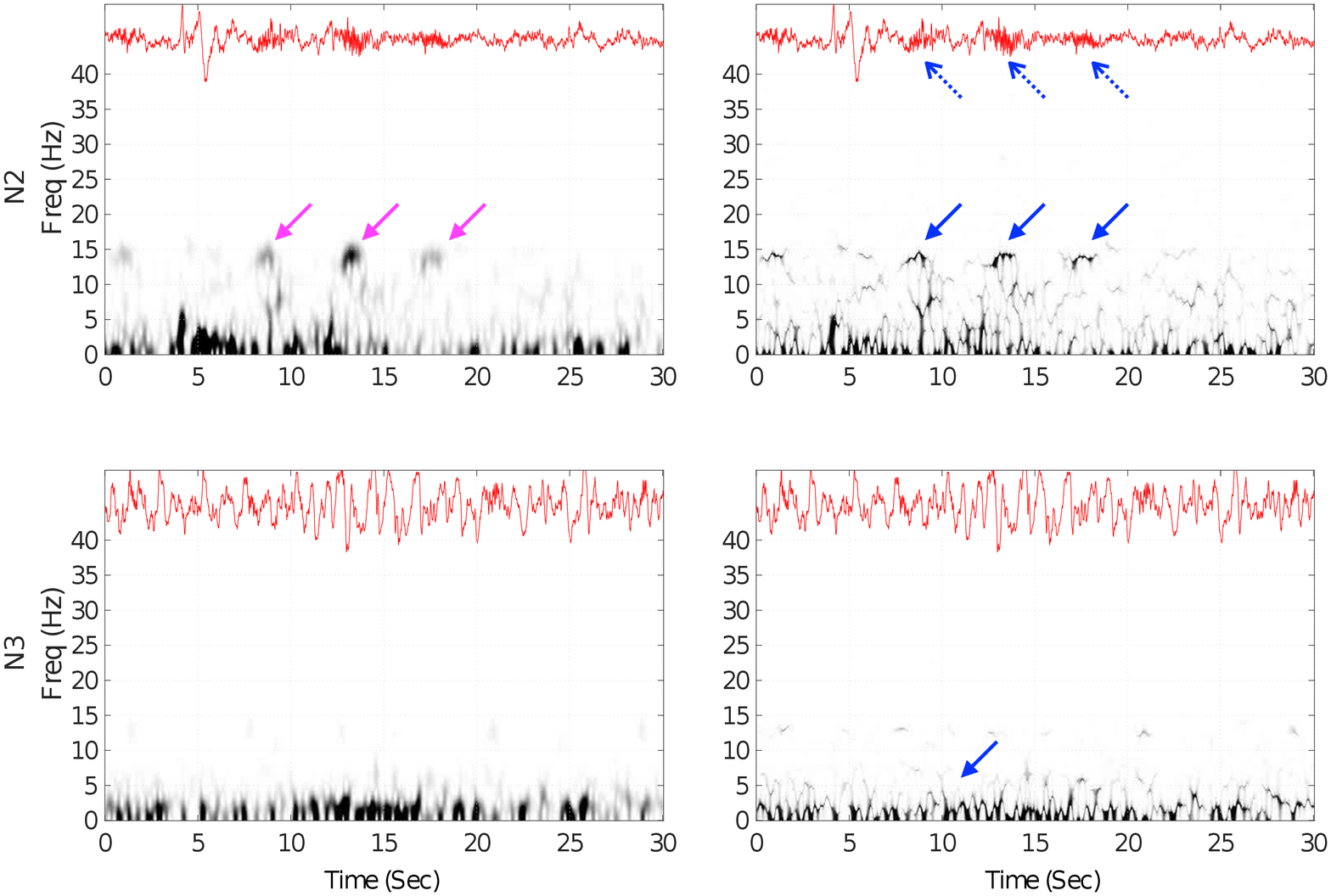}
\caption{An illustration of EEG spectrogram (left) and synchrosqueezed EEG spectrogram (right) of sleep stages N2 and N3 from Sleep-EDF SC$^*$ database. The EEG signals are superimposed as red curves for visual comparison.
}
\label{Figure1}
\end{figure}

To better appreciate the effect of the reassignment step, see Figures \ref{Figure0} and \ref{Figure1} for a comparison of the synchrosqueezed spectrogram and the spectrogram of an EEG signal during different sleep stages. We call the spectrogram and synchrosqueezed spectrogram of an EEG signal {\em EEG spectrogram} and {\em synchrosqueezed EEG spectrogram} respectively.
In these figures, the EEG signal is superimposed as red curves for a visual comparison.
Compared with the EEG spectrogram, the synchrosqueezed EEG spectrogram is sharper since the phase information is taken into account for the reassignment; for example, the alpha wave from 13-th second to 21-th second of REM that oscillates at about 10 Hz {(blue arrows in the middle right plot in Figure \ref{Figure0})} and the spindles around 8-th, 13-th and 17-th seconds of N2 that oscillates at about 14 Hz {(blue arrows in the top right plot in Figure \ref{Figure1})} can be clearly visualized in the synchrosqueezed EEG spectrogram {(magenta arrows in the middle left plot in Figure \ref{Figure0} and magenta arrows in the top left plot in Figure \ref{Figure1})}.
{These oscillation behavior can also be found in the EEG signal (blue dashed arrows in the middle right plot in Figure \ref{Figure0} and blue dashed arrows in the top right plot in Figure \ref{Figure1}), but harder to quantify. In the EEG spectrogram, these curves are blurred and not easy to directly identity the variation of the time-varying frequency. Quantifying the time-vary frequency is critical for further understanding the physiology of sleep dynamics, and its study will be reported in the future work.}
Although the theta wave {(blue arrows in the bottom right plot in Figure \ref{Figure0}) and delta wave (blue arrows in the bottom right plot in Figure \ref{Figure1})} in N1 and N3 stages have less regular oscillatory pattern, the synchrosqueezed EEG spectrogram is more concentrated.

To sum up, the SST-based approach takes the phase information into account and helps avoid energy leakage due to uncertainty principle. As a result, it prevents the possibility of failing to sum up all the signal energy corresponding to one mode centered in a given subband, in the case that the STFT energy leaks into another subband.

\section{Local Mahalanobis distance and empirical intrinsic geometry}\label{Section:MetricDesignEIG}

To recover the intrinsic sleep dynamics, we need a sophisticated metric to help organize features obtained from SST. In this paper, we consider the {\em local Mahalanobis distance} as the metric. To motive how the local Mahalanobis distance is designed and how it is incorporated into the inter-individual prediction framework, below we review the dynamics system model, the empirical intrinsic geometry (EIG) model \cite{singer2008non,TalmonPNAS}. Then we show how to generalize the EIG model and detail the desired local Mahalanobis distance.

We start from recalling the EIG model as the motivation. %The EIG model is designed specifically for a single subject. For each subject, w
We assume that the point cloud, or features, denoted as $\{\mathbf{u}^{(j)}\}_{j=1}^n\subset \mathbb{R}^{q}$, comes from a diffeomorphic deformation $\Phi: \mathbb{R}^{p}\rightarrow \mathbb{R}^{q}$ which maps the latent intrinsic state space that hosts the dynamical system $\bm \theta^{(j)}$ that describes the dynamics we have interest; that is, $\mathbf{u}^{(j)}=\Phi(\bm \theta^{(j)})$, where $p>0$ is the dimension of the inaccessible dynamical space. We call $\Phi$ the {\em observation transform}.
Furthermore, assume that the inaccessible intrinsic state $\bm \theta^{(j)}$ is the value of the process $\bm \theta$ at the $j$-th sampling time stamp and $\bm \theta$ satisfies the stochastic differential equation
\begin{equation}\label{SDE_theta}
d\bm\theta(t)=a(\bm\theta(t))dt+d\boldsymbol\omega(t)
\end{equation}
where $a$ is an unknown drifting function and $\boldsymbol\omega$ is the standard $d$-dim Brownian motion. This latent space model is called EIG in \cite{singer2008non,TalmonPNAS}, and it has been widely considered to designed other algorithms, like Kalman filter.
Based on (\ref{SDE_theta}), it is shown in \cite[Section 3]{singer2008non} that when the intrinsic distance of $\mathbf{u}^{(i)}$ and $\mathbf{u}^{(j)}$ is sufficiently small, we could recover the intrinsic distance between {$\bm{\theta}^{(i)}$ and $\bm{\theta}^{(j)}$} by
\begin{equation}\label{EIG_dis}
\|\bm{\theta}^{(i)}-\bm{\theta}^{(j)}\|_{\mathbb{R}^{d}}^{2}=\frac{1}{2}\big(\mathbf{u}^{(i)}-\mathbf{u}^{(j)}\big)^{\top}
\big[(\mathbf{C}^{(i)})^{-1}+(\mathbf{C}^{(j)})^{-1}\big]\big(\mathbf{u}^{(i)}-\mathbf{u}^{(j)}\big),
\end{equation}
where $\mathbf{C}^{(i)}=\nabla \Phi(\bm{\theta}^{(i)})[\nabla \Phi(\bm{\theta}^{(i)})]^{\top}$ is the covariance matrix associated with the deformed Brownian motion, up to the error term $O(\|\mathbf{u}^{(i)}-\mathbf{u}^{(j)}\|^{4})$. Furthermore, it is shown in \cite{singer2008non,TalmonPNAS} that $\mathbf{C}^{(i)}$ can be estimated by the covariance matrix of $\{\mathbf{u}^{(k)}\}_{k=i-\delta}^{i+\delta}$, where $\delta$ is a predetermined integer.
This face comes from the Ito's formula. Note that by the Ito's formula, we immediately have
\begin{equation}
\mathrm{d}\mathbf{u}_{t}=\left(\frac{1}{2}\Delta\Phi|_{\bm\theta_{t}}+\nabla\Phi|_{\bm\theta_{t}}a(\bm\theta_{t})\right)\mathrm{d}t+\nabla\Phi|_{\theta_{t}}\mathrm{d}\omega_{t}.
\end{equation}
Since $(\frac{1}{2}\Delta\Phi|_{\bm\theta_{t}}+\nabla\Phi|_{\bm\theta_{t}}a(\bm\theta_{t}))\mathrm{d}t$ is the drifting term, we know
\begin{equation}
\text{Cov}\left(\mathrm{d}\mathbf{u}_{t}\right)=\nabla\Phi|_{\bm\theta_{t}}\nabla\Phi|_{\bm\theta_{t}}^\top\,.
\end{equation}
We call $\|\bm{\theta}^{(i)}-\bm{\theta}^{(j)}\|_{\mathbb{R}^{d}}$ the {\em local Mahalanobis distance} between $\mathbf{u}^{(i)}$ and $\mathbf{u}^{(j)}$, which is a generalization of Mahalanobis distance from the statistical viewpoint.

While the EIG model works well for a single subject \cite{Katz_Wu_Lo_Talmon:2016}, it may not be suitable for the inter-subject sleep assessment mission {in which} we have interest. Indeed, since different subjects have different sleep dynamics, and the observation transforms are different, physiologically it is not reasonable to quantify the intrinsic state dynamics of different {subjects} by a single equation like (\ref{SDE_theta}). Indeed, it does not make sense from the physiological viewpoint to integrate the temporal sleep dynamics of different subjects into one; that is, there is no temporal relationship between epochs from different subjects.
In this work, motivated by the success of (\ref{SDE_theta}), we consider the following generalization of the EIG model to study the synchrosqueezed EEG spectral features.
Based on the physiological assumption that the EEG signals of different subjects share similar spectral behavior, we assume that the synchrosqueezed EEG spectral features from different subjects {come} from the same map $\Psi: \mathbb{R}^{p}\rightarrow \mathbb{R}^{10}$, which maps the inaccessible intrinsic state $\bm \theta^{(j)}$, where $0<d\leq p$ is the dimension of the inaccessible space hosting the state space of the dynamics, which is assumed to be the same among subjects, to the space hosting the synchrosqueezed EEG spectral features. We further assume that $p=10$ and $\Psi$ is an identity from the state space to its range perturbed by a stationary random perturbation that has mean 0 and the covariance $I_{p\times p}$. To simplify the terminology, we still call $\Psi$ the {\em observation transform}.
Note that this is the simplified EIG model with noise in the observation considered in \cite{Talmon2013}.

Under this simplified EIG model, we can get an estimate for $\|\bm{\theta}^{(i)}-\bm{\theta}^{(j)}\|_{\mathbb{R}^{p}}^{2}$ similar to (\ref{EIG_dis}) by modifying the proof of \cite{singer2008non}.
Denote the $K$-neighborhood of $\mathbf{u}^{(j)}$ by $\mathcal{N}_j$ for each $j\in\{1,2,\ldots,n\}$.
Based on the assumption of $\Psi$, when the data is supported on a $d$-dimensional smooth manifold embedded in $\mathbb{R}^p$, we have
\begin{equation}\label{cov1}
\|\bm{\theta}^{(i)}-\bm{\theta}^{(j)}\|^{2}_{\mathbb{R}^{p}}\approx
(\mathbf{u}^{(i)}-\mathbf{u}^{(j)})^{\top}\mathcal{T}_d(\nabla \Psi|_{\bm{\theta}^{(j)}}
\nabla \Psi|_{\bm{\theta}^{(j)}}^{\top})
(\mathbf{u}^{(i)}-\mathbf{u}^{(j)}),
\end{equation}
where $\mathcal{T}_d(\nabla \Psi|_{\bm{\theta}^{(j)}}\nabla \Psi|_{\bm{\theta}^{(j)}}^{\top})$ means the truncated pseudo-inverse of $\nabla \Psi|_{\bm{\theta}^{(j)}}\nabla \Psi|_{\bm{\theta}^{(j)}}^{\top}$ defined by $U\Lambda^\dagger_dU^\top$, $\nabla \Psi|_{\bm{\theta}^{(j)}}\nabla \Psi|_{\bm{\theta}^{(j)}}^{\top}=U\Lambda U^\top$ is the eigendecomposition, $\Lambda^\dagger_d=\textup{diag}[\ell_1^{-1},\ldots,\ell_d^{-1},0,\ldots,0]$, and $\Lambda=\textup{diag}[\ell_1,\ldots,\ell_p]$ with $\ell_1\geq \ell_2\geq\ldots\geq \ell_p\geq 0$.
Similarly, the covariance of the stationary random perturbation associated with $\Psi$ becomes
\begin{align}
\Gamma_{j}:=\,&\frac{1}{K}\underset{i\in \mathcal{N}_{j}}{\sum}(\mathbf{u}^{(i)}-\mathbf{u}^{(j)}) (\mathbf{u}^{(i)}-\mathbf{u}^{(j)})^{\top}\label{cov2}
\\\notag=\,&\frac{1}{K}\underset{i\in \mathcal{N}_{j}}{\sum} \nabla \Psi|_{\bm{\theta}^{(j)}}(\bm{\theta}^{(i)}-\bm{\theta}^{(j)})(\bm{\theta}^{(i)}-\bm{\theta}^{(j)})^{\top}[\nabla \Psi|_{\bm{\theta}^{(j)}}]^{\top}\nonumber\\
\approx\,&
\nabla \Psi|_{\bm{\theta}^{(j)}}[\nabla \Psi|_{\bm{\theta}^{(j)}}]^{\top}.\nonumber
\end{align}
Combining (\ref{cov1}) and (\ref{cov2}) yields
\begin{equation}\label{cov3}
\|\bm{\theta}^{(i)}-\bm{\theta}^{(j)}\|^{2}_{\mathbb{R}^{d}}\approx
(\mathbf{u}^{(i)}-\mathbf{u}^{(j)})^{\top}\mathcal{T}_d[\Gamma_{j}]
(\mathbf{u}^{(i)}-\mathbf{u}^{(j)}).\nonumber
\end{equation}
Therefore, following the observation in (\ref{EIG_dis}), we thus consider the following metric:
\begin{equation}\label{Definition:EIG_dis2}
d_{\texttt{LMD}}(\mathbf{u}^{(i)},\mathbf{u}^{(j)})^{2}=
\frac{1}{2}(\mathbf{u}^{(i)}-\mathbf{u}^{(j)})^{\top}(\mathcal{T}_d[\Gamma_{i}]+\mathcal{T}_d[\Gamma_{j}] )
(\mathbf{u}^{(i)}-\mathbf{u}^{(j)})\,,
\end{equation}
which we call the {\em local Mahalanobis distance} (MD).
We mention that this approach leads to a more accurate geodesic distance estimation by a direct hard threshold of the noisy covariance matrix to remove the influence of noise. A more general discussion can be found in \cite{JohnNan2018}.

\section{Diffusion map}\label{sec:DM}

Graph Laplacian (GL) based algorithms have attracted a lot of attention in the machine learning society. DM \cite{coifman2006diffusion} is one of those successful algorithms. To better understand the theoretical foundation of DM, in the past decade, several works have been done based on the differential geometry framework. {The behavior of spectral embedding under the spectral geometry is discussed in \cite{Berard_Besson_Gallot:1994}.
Later, based on the manifold model, how the GL converges to the Laplace-Beltrami operator is studied in \cite{Belkin_Niyogi:2005,Hein_Audibert_Luxburg:2005,Singer:2006}.
The spectral convergence of GL is studied in \cite{Singer_Wu:2016}. The spectral convergent rate is reported in \cite{trillos2018error,Wang:2015}. In \cite{Gine_Koltchinskii:2006} the central limit theory of GL is provided.
The problem of embedding by finite eigenfunctions of Laplace-Beltrami operator is studied in \cite{Jones2008,Bates:2014,Portegies:2015}. Below we summarize those theoretical results.}
We start from recalling the DM algorithm. Given a dataset $\mathcal{X}:=\{x_i\}$. Construct a $n\times n$ affinity matrix $W$ so that
\begin{align}\label{Definition W matrix}
W_{ij}=e^{-d(x_i,\,x_j)^2/\epsilon},\ \ \mbox{for}\ \ i,j=1,\ldots,n,
\end{align}
where $d(\cdot,\cdot)$ is the chosen metric and the bandwidth $\epsilon>0$ is chosen by the user\footnote{According to the noise analysis in \cite{ElKaroui_Wu:2016b}, when the signal to noise ratio is small, it is beneficial to set the diagonal terms of the affinity matrix to $0$; that is, set $W_{ii}=0$.}. The affinity is clearly the composition of the radial basis function kernel $K(t)=e^{-t^2/\epsilon}$ and the distance $d(x_i,\,x_j)$. In general, we can choose more general kernels with a sufficient decay rate. To simplify the discussion, we focus on the radial basis function kernel.
Next, define a diagonal matrix $D$ of size $n\times n$ as
\begin{equation}\label{degree_matrix}
D(i,i)=\sum_{j=1}^nW(i,j),\ \ \textup{for}\ \ i=1,\ldots,n.
\end{equation}
In general, $D$ is called the degree matrix.
With matrices $W$ and $D$, define a random walk on the point cloud $\mathcal{X}$ with the transition matrix given by the formula
\begin{align}\label{Definition:Atransition}
A:=D^{-1}W\,.
\end{align}
Clearly, $A$ is diagonalizable since $A$ is similar to the symmetric matrix $D^{-1/2}WD^{-1/2}$. Therefore, it has a complete set of right eigenvectors $\phi_1, \phi_2,\cdots,\phi_n\in \mathbb{R}^n$ with corresponding eigenvalues $1 = \lambda_1 > \lambda_2\geq\cdots\geq\lambda_n\geq 0$, where $\phi_1 = [1, 1, \ldots , 1]^\top\in \mathbb{R}^n$. Indeed, from the eigen-decomposition $D^{-1/2}WD^{-1/2}=O\Lambda O^\top$, where $O\in O(n)$ and $\Lambda=\text{diag}[\lambda_1,\ldots,\lambda_n]$ is a $n\times n$ diagonal matrix, we have $A=U\Lambda V^\top$, where $U=D^{-1/2}O$ and $V=D^{1/2}O$.
Note that $\lambda_1>\lambda_2$ since we assume the graph is complete, and hence connected, and $\lambda_n\geq 0$ comes from the chosen kernel and the Bochner theorem \cite{Gelfand:1964}.
With the decomposition $A=U\Lambda V^\top$, the DM is defined as
\begin{equation}\label{DM}
\Phi_t:x_j \mapsto \big(\lambda_2^t\phi_2(j), \lambda_3^t\phi_3(j),\ldots,\lambda_{\hat{d}+1}^t\phi_{\hat{d}+1}(j)\big)\in \mathbb{R}^{\hat{d}}\,,
\end{equation}
where $j=1,\ldots,n$, $t>0$ is the diffusion time chosen by the user, and $\hat{d}$ is the embedding dimension chosen by the user. Note that $\lambda_1$ and $\phi_1$ are ignored in the embedding since they are not informative. In practice, in addition to determining $\hat{d}$ by a direct assignment, $\hat{d}$ can be determined by a more adaptive way according to the decay of the eigenvalue decay; for example, $\hat{d}$ can be chosen to be the largest $j$ so that $\lambda_{j}^t>\delta>0$, where $\delta$ is chosen by the user. Both can be obtained by optimizing some quantities of interest based on the problem at hand.
Clearly, $\Phi_t(x_j)$ consists of the second to $(\hat{d}+1)$-th coordinates of $e_j^\top U\Lambda^t$, where $e_j$ is the unit $n$-dim vector with the $j$-th entry $1$. With the DM, we can define the {\em diffusion distance} (DD) with the diffusion time $t>0$ between $x_i$ and $x_j$ as
\begin{equation}\label{Definition:DD}
D_t(x_i,x_j)=\|\Phi_t(x_i)-\Phi_t(x_j)\|_{\mathbb{R}^{\hat{d}}}\,.
\end{equation}

We now summarize the theory behind DM under the manifold model. Supposing a data set $\mathcal{X}=\{x_i\}_{i=1}^n\subset \mathbb{R}^p$ is independently and identically sampled from a random vector $X:(\Omega,\mathcal{F},\mathbb{P})\to \mathbb{R}^p$, where the range of $X$ is supported on a low dimensional manifold $M$ embedded in $\mathbb{R}^p$. We assume that the manifold is compact and smooth and the metric $g$ is induced from $\mathbb{R}^p$. We assume that the induced measure on the Borel sigma algebra on $M$, denoted as $X_*\mathbb{P}$, is absolutely continuous with respect to the Riemannian measure $dV_g$. Furthermore, we assume that the function $p=\frac{dX_*\mathbb{P}}{dV_g}$ by Radon-Nykodin theorem is bounded away from zero and is sufficiently smooth.
When $n\to\infty$, the transition matrix $A$ defined in (\ref{Definition:Atransition}) converges to a continuous diffusion operator defined on the manifold when the metric in \eqref{Definition W matrix} is chosen to be $d(x_i,x_j)=\|x_i-x_j\|_{\mathbb{R}^p}$ \cite{Belkin_Niyogi:2005,Hein_Audibert_Luxburg:2005,Singer:2006}. Under the smooth manifold setup, the geodesic distance between two close points can be well approximated by the Euclidean distance; see, for example, \cite[Lemma 2,Theorem 2]{JohnNan2018}.
If the sampling scheme is uniform, we can further approximate the Laplace-Beltrami operator of the manifold, denoted as $\Delta_g$, when $\epsilon \to 0$; if the sampling scheme is non-uniform but smooth enough, by estimating the density function we could correct the diffusion process by the {\em $\alpha$-normalization} scheme proposed in \cite{coifman2006diffusion}, and again approximate the Laplace-Beltrami operator of the manifold when $\epsilon \to 0$. The convergence happens both in the pointwise sense and spectral sense \cite{trillos2018error,Wang:2015,Singer_Wu:2016}. In summary, we can view the eigenvectors and eigenvalues of $\frac{A-I}{\epsilon}$ as approximation of the eigenfunctions and eigenvalues of the Laplace-Beltrami operator associated with the manifold.

With the Laplace-Beltrami operator, we could apply the spectral embedding theory \cite{Berard:1986,Berard_Besson_Gallot:1994} to embed the manifold (and hence the data) using the eigenfunctions of the diffusion operator.
Suppose the manifold is connected, and the $l$-th eigenvalue of $\Delta_g$ is $-\mu_l$ with the eigenfunction $f_l$; that is, $ f_l=-\mu_lf_l$, where $\mu_1=0<\mu_2\leq \mu_3,\ldots$. Note that since the manifold is connected, $\mu_2>0$, and when $\mu_l$ has a non-trivial multiplicity, $f_l$ might not be unique. With a given set of eigenvalues and eigenvectors $\{(\mu_l,f_l)\}_{l=1}^\infty$, the {\em spectral embedding} is defined as \cite{Berard_Besson_Gallot:1994}
\begin{equation}
\phi_t:x\mapsto c(e^{-t\mu_l}f_l(x))_{l=2}^\infty\in \ell_2\,,
\end{equation}
where $t>0$ is the {\em diffusion time}, and $c$ is the normalization constant depending on $t$. Note that the embedding defined in (\ref{DM}) is a discretization of $\phi_t$. Indeed, as is shown in \cite[Theorem 5.4]{Singer_Wu:2016}, for $t>0$, $A^{t/\epsilon}$ will converge to the heat kernel of $e^{-t\Delta_g}$ in the spectral sense when $n\to \infty$ and $\epsilon=\epsilon(n)\to 0$ satisfying some mild conditions. Therefore, $\lambda_l$ in (\ref{DM}) converges to $\mu_l$, and $\phi_l$ in (\ref{DM}) converges to $f_l$. \footnote{Note that $\phi_l$ is a $n$-dim vector while $f_l$ is a smooth function defined on $M$. To properly state the convergence, we need to convert $\phi_l$ into a continuous function defined on $M$. Also, when $\mu_l$ has a non-trivial multiplicity, the  convergence should be stated using the eigenprojection operator. We refer these technical details to \cite{Singer_Wu:2016}}
We also define $d_t(x,y)=\|\phi_t(x)-\phi_t(y)\|_{\ell_2}$ to be the diffusion distance. Again, \eqref{Definition:DD} is a discretization of $d_t(x,y)$.
This embedding allows us to reveal the geometric and topological structure of the manifold, and hence the structure of the dataset.
In particular, it is shown in \cite{Berard_Besson_Gallot:1994} that $\phi_t$ is an almost isometric embedding when $t$ is small enough; in \cite{Singer_Wu:2012}, it is shown that the local geodesic distance of the manifold could be well approximated by the diffusion distance, when the diffusion time is long enough compared with the geodesic distance. Lastly, when combined with the finite dimensional embedding result of the spectral embedding theory of the Laplace-Beltrami operator shown in \cite{Bates:2014,Portegies:2015}, we could guarantee that with finite sampling points, when $n$ is large enough, we can reconstruct the manifold with a given accuracy. This is the {\em embedding property} that we expect to reorganize the dataset.

The ability to reconstruct the underlying intrinsic structure is not the only significant strength of DM. It has been shown in \cite{ElKaroui:2010a,ElKaroui_Wu:2016b} that DM is also robust to noise in the following sense. Suppose the data point $y_i\in \mathcal{Y}$ comes from contaminating a clean sample $x_i$ by {\em some} noise $\xi_i$. Suppose $\xi_i$ satisfies some mild conditions, like {finite variance and reasonably controlled noise level}. Note that we do not require $\xi_i$ to be identical from point to point. Denote the transition matrix built up from $\{y_i\}_{i=1}^n$ as $W^{(\texttt{noisy})}$. Under this condition, the deviation of $W^{(\texttt{noisy})}$ from $W$ is well controlled by the noise level in the norm sense. Thus, we conclude that the eigenvectors of {$W$} with sufficiently large eigenvalues could be well reconstructed from {$W^{(\texttt{noisy})}$}. This is the {\em robustness property} we expect from DM to analyze the noisy data.

With the embedding property and the robustness property, we can well approximate the underlying geometric structure from the noisy dataset. To show this, recall that the clean points $\{x_i\}_{i=1}^n$ are sampled from a manifold as discussed above.
By Weyl's law \cite{Berard:1986}, and the spectral convergence of DM, the eigenvalues of $W_0$ decay exponentially fast. Therefore, by the robustness property, the first few eigenvectors and eigenvalues could be well reconstructed. However, the eigenvectors with small eigenvalues will be highly contaminated. Since eigenvalues of the clean data decay fast, those eigenvectors with small eigenvalues are not that informative from the spectral embedding viewpoint -- although DM is a nonlinear map, when $\hat{d}$ is chosen large enough, the finite dimensional embedding result guarantees that we can reconstruct the manifold, and hence DM is an almost isometric embedding. Therefore, we conclude that DM with the truncation scheme allows us to a well reconstruction of the clean data up to a tolerable error. We call this property the {\em recovery property}.

\section{Sensor fusion by alternating diffusion, co-clustering and multiview DM}\label{Section:Multiview}

When we have two sensors collecting data simultaneously from the system of interest, a common question to ask is if we can integrate information from both sensors and achieve a better representation/feature for the system. This problem is understood as the {\em sensor fusion} problem. In our motivating sleep dynamics problem, while different EEG channels capture information from the same brain, the information recorded might vary and they might be contaminated by brain-activity irrelevant artifacts from different {sources}, including noise and other sensor-specific nuisance. These artifacts not only deteriorate the quality of the extracted features but might also mislead the analysis result. The main purpose of sensor fusion is distilling the brain information and removing those unwanted artifacts.

To {simplify} the discussion, we assume that we have two simultaneously recorded datasets $\mathcal{X}=\{x_i\}_{i=1}^n$ and $\mathcal{Y}=\{y_i\}_{i=1}^n$ (for example, two EEG channels); that is, $x_i$ and $y_i$ are recorded at the same time. While in general $x_i$ and $y_i$ can be of complicated data format, to simplify the discussion, assume $\mathcal{X}\subset \mathbb{R}^{d_x}$ and $\mathcal{Y}\subset \mathbb{R}^{d_y}$ respectively. In other words, we may view $x_i$ and $y_i$ as the $i$-th features captured by two sensors. A naive way to combine information from $\mathcal X$ and $\mathcal Y$ is simply concatenating $x_i$ and $y_i$ and form a new feature of size ${d}_x+{d}_y$.
However, due to the inevitable sensor-specific artifacts or errors, such a concatenating scheme might not be the optimal route to fuse sensors \cite{lederman2015alternating,Talmon_Wu:2016}.
We consider the recently developed diffusion-based approaches to fuse sensor information, including AD \cite{lederman2015alternating,Talmon_Wu:2016} and co-clustering \cite{dhillon2001co} (a special case of multiview DM \cite{Linderbaum_Yeredor_Salhov_Averbuch:2015}).
In short, the information from different sensors are ``diffused'' to integrate the {nonlinear} common information shared between different sensors, and simultaneously eliminate artifacts or noise specific to each sensor. {This is a nonlinear generalization of the well known canonical correlation analysis (CCA) \cite{Hardle2007}.} While we focus on the diffusion-based approaches in this work, there is a huge literature about sensor fusion, and we refer readers to \cite{lahat2015multimodal} for a systematic review.

\subsection{Alternating diffusion}\label{Subsection:AD}
For AD, we first form two transition matrices $A_x\in \mathbb{R}^{n\times n}$ and $A_y\in \mathbb{R}^{n\times n}$ from $\mathcal X$ and $\mathcal Y$ respectively following the definition \eqref{Definition:Atransition}. Then, form a new transition matrix:
\begin{equation}\label{ADM_matrix}
A_{xy}=A_{x}A_{y}\in \mathbb{R}^{n\times n}\,.
\end{equation}
Note that $A_{xy}\boldsymbol 1=A_xA_y\boldsymbol 1=A_x\boldsymbol 1$. Thus, $A_{xy}$ can be viewed as a transition matrix associated with the random walk on the ``joint data set'' $\{z_i\}_{i=1}^n$, where $z_i=(x_i,y_i)$, while the associated affinity graph is directed. In fact, if we write $A_x=D_x^{-1}W_x$ and $A_y=D_y^{-1}W_y$, we have
\[
A_{xy}=D_x^{-1}(W_xD_y^{-1}W_y).
\]
Note that $W_xD_y^{-1}W_y$ is a non-negative matrix and is in general non-symmetric. If we view $W_xD_y^{-1}W_y$ as an affinity matrix, the associated degree matrix is $D_x$. Thus, the affinity graph associated with $W_xD_y^{-1}W_y$ is directed.
On the other hand, $A_{xy}$ can be viewed as starting a random walk on $\mathcal Y$, jumping to $\mathcal X$, and continuing another random walk on $\mathcal X$. This is the motivation of the terminology AD. Clearly, AD depends on the order of multiplication.
In \cite{lederman2015alternating}, it is proposed that we determine the intrinsic distance between the $i$-th sample pair, $(x_i,y_i)$, and the $j$-th sample pair, $(x_j,y_j)$, by the $\ell^2$ norm of the $i$-th row of $A_{xy}$ and the $j$-th row of $A_{xy}$. With this intrinsic distance, it is suggested in \cite{lederman2015alternating} to apply another DM.

To proceed, we need the ``spectral decomposition'' of $A_{xy}$. When $A_{xy}$ is diagonalizable, we apply the spectral decomposition to get $A_{xy}=\Phi_{xy}\Lambda_{xy} \Psi_{xy}^{-1}$, where $\Phi_{xy},\,\Psi_{xy}\in Gl(n)$. Then we may proceed with the {\em alternating diffusion map} (ADM) \cite{Talmon_Wu:2016} by embedding the $j$-th sample to an Euclidean space:
\[
\Phi_{xy,t}:(x_j,y_j)\mapsto \big(\lambda_{xy,2}^t\phi_{xy,2}(j), \lambda_{xy,3}^t\phi_{xy,3}(j),\ldots,\lambda_{xy,d_{xy}+1}^t\phi_{xy,d_{xy}+1}(j)\big)\in \mathbb{R}^{d_{xy}},
\]
where $\lambda_{xy,l}$ and $\phi_{xy,l}$ is the $l$-th right eigenpair, and $d_{xy}$ is the embedding dimension chosen by the user. The discussion of the embedding is the same as that of DM. However, in general $A_{xy}$ is not diagonalizable, even if $A_x$ and $A_y$ are both diagonalizable. In this case, we may consider the singular value decomposition (SVD) of $A_{xy}$, and proceed with the embedding by taking the singular value and singular vector into account \cite{Talmon_Wu:2016,marshall2018time}.
We mention that AD is closely related to a nonlinear generalization of CCA, called {\em nonlinear CCA} (NCCA) \cite{michaeli2016nonparametric}. In NCCA, when the kernel is chosen to be Gaussian, the main operator of interest is the SVD of
\[
C_{xy}:=A_xA_y^\top.
\]
Unlike AD, $C_{xy}$ is in general not a transition matrix. However, it is related to the mutual information between $\mathcal{X}$ and $\mathcal{Y}$. We refer readers with interest to \cite{michaeli2016nonparametric} for more details. Note that both $A_{xy}$ and $C_{xy}$ are in general not diagonalizable and depend on the order of multiplication. To remedy this drawback, in \cite{shnitzer2018recovering}, a symmetrized AD is proposed, that is,
\[
A=A_{x}A_{y}^\top +A_{y}A_{x}^\top \in \mathbb{R}^{n\times n}.
\]
Again, $A$ is in general not a transition matrix. We refer readers with interest to \cite{shnitzer2018recovering} for more details.
For more discussion about AD, ADM and NCCA, we refer the reader with interest to \cite{lederman2015alternating,Talmon_Wu:2016,michaeli2016nonparametric}.
When there are more than two channels, a possible generalization of the following discussion can be found in \cite{Katz_Wu_Lo_Talmon:2016} or \cite[(28)]{lederman2015alternating}.

\begin{figure}
\centering
\includegraphics[scale=0.47]{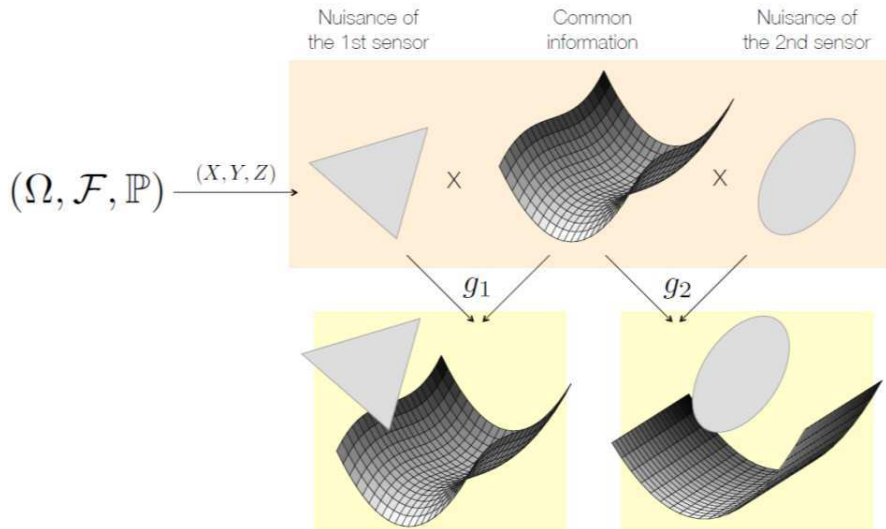}%{ADmodeleps.eps}
\caption{Illustration of the mathematical model for the sensor fusion algorithm, alternating diffusion.
}
\label{FigureADmodel}
\end{figure}

We now summarize some theoretical analyses of AD. Consider three latent random variables $X$, $Y$, and $Z$ and assume that the associated joint probability density satisfies
\[
f_{X,Y,Z}(x,y,z)=f_X(x)f_{Y|X}(y|x)f_{Z|X}(z|x)\,,
\]
that is, $Y$ and $Z$ are conditionally independent given $X$. The data collected from the first sensor is modeled as
\[
S^{(1)}=g_1(X,Y),
\]
and that of the second sensor is modeled as
\[
S^{(2)}=g_2(X,Z),
\]
where we assume $g_1$ and $g_2$ satisfy some regularity; for example, $g_1$ and $g_2$ are both bilipschitz. In this model, $X$ is the system we have interest to study, while $Y$ and $Z$ are {\em noises} (or artifacts, nuisances) associated with the first and second sensor respectively. We call $X$ the {\em common information}. When the first (resp. second) sensor collects data from the system $X$, the sensor-dependent noise $Y$ (resp. $Z$) comes into play via $g_1$ (resp. $g_2$). Specifically, the sample datasets $\{s^{(i)}_{j}\}_{j=1}^n$, where $i=1,2$, come from mapping $(x_i,y_i,z_i)$ i.i.d. sampled from the latent space $(X,Y,Z)$ via $g_i$; that is, $s^{(1)}_j=g_1(x_j,y_j)$ and $s^{(2)}_j=g_2(x_j,z_j)$, $j=1,\ldots,n$. See Figure \ref{FigureADmodel} for an illustration of the model, where $(\Omega, \mathcal{F},\mathbb{P})$ means the event space $\Omega$, the sigma algebra $\mathcal{F}$ and the probability measure $\mathbb{P}$. Note that in general, the common information might be deformed via the observation process, $g_1$ or $g_2$. The same model can be easily generalized to the case when there are multiple sensors

Following the analysis in \cite{lederman2015alternating}, the $\ell^2$ distance between the $i$-th row and the $j$-th row of $A_{xy}$ serves as a good estimate of the distance on the common information associated with the $i$-th and the $j$-th samples. We call this distance the {\em common metric}. In other words, although two sensors are contaminated by different noises, AD allows us a stable recovery of the common information. In \cite{Talmon_Wu:2016}, it is shown that when the common information can be modeled as a Riemannian manifold, AD recovers the Laplace-Beltrami operator of the manifold when $g_1$ and $g_2$ do not deform the common information. If either $g_1$ or $g_2$ deform the common information, under mild assumption about the deformation, AD leads to a ``deformed'' Laplace-Beltrami operator. In other words, it says that although $A_{xy}$ is in general not symmetric, its asymptotic is a self-adjoint operator. We refer readers to \cite{Talmon_Wu:2016} for technical details.
The behavior of ADM can be analyzed by combining the recovery property based on spectral geometry summarized in Section \ref{sec:DM}. More results about symmetrized AD can be found in \cite{shnitzer2018recovering}.

\subsection{Co-clustering and multiview DM}

We now discuss the well known co-clustering algorithm \cite{dhillon2001co} based on the bipartite graph model.
Denote ${V}_{1}=\mathcal X=\{x_{i}\}_{i=1}^n$ (resp. ${V}_{2}=\mathcal Y=\{y_{i}\}_{i=1}^n$) to be the set of $n$ vertices representing the $n$ feature vectors extracted from the first (resp. second) sensor.
Form a bipartite graph $G=(\mathbb{V},\mathbb{E})$, where $\mathbb{V}$ consists of $2n$ vertices from $V_{1}$ and $V_{2}$ and $\mathbb{E}$ is the set of edges between vertices in $V_{1}$ and vertices in $V_{2}$; that is, edges only exist between $V_{1}$ and $V_{2}$ and there is no edge inside $V_{1}$ and there is no edge inside $V_{2}$.
Then, by assigning affinities on all edges, $(x_i, y_{j})$ for all $i,j=1,\ldots,n$, we obtain a bipartite affinity graph. The affinity is determined by the user.
Denote
\[
\mathsf M\in \mathbb{R}^{2n\times 2n}
\]
to be the affinity matrix associated with the constructed bipartite affinity graph, where the first $n$ columns and rows of $\mathsf M$ are associated with the first sensor, and the last $n$ columns and rows are associated with the second sensor.

In \cite{dhillon2001co}, based on $\mathsf M$, it is argued how we can determine the corresponding clusters in ${V}_{2}$ based on the clusters in ${V}_{1}$, and vice versa.
More precisely, given disjoint clusters ${V}_{1,1},\ldots,{V}_{1,K}$ for vertices in ${V}_{1}$, where $K$ is a predetermined integer, the clusters ${V}_{2,1},\ldots,{V}_{2,K}$ for vertices in ${V}_{2}$
can be formed by
\begin{equation}\label{equ:multiview4}
{V}_{2,m} = \Big\{w_{j}\in {V}_{2}:\ \sum_{x_i\in {V}_{1,m}} \mathsf M_{i,j}\geq \underset{x_i\in {V}_{1,\ell}}{\sum}\mathsf M_{i,j}\ \textup{for any}\ \ell=1,\ldots,K \Big\},
\end{equation}
where $m=1,\ldots,K$. The motivation beyond this assignment is intuitive -- a given recording $x_i$ from the second channel has a higher chance to belong to the $j$-th cluster if it more likely belongs to the $j$-th cluster of the first channel than other clusters of the first channel.
Similarly,
given $K$ disjoint clusters ${V}_{2,1},\ldots,{V}_{2,K}$
for vertices in ${V}_{2}$, the induced clusters ${V}_{1,1},\ldots,{V}_{1,K}$ for vertices in ${V}_{1}$
are determined by
\begin{equation}\label{equ:multiview5}
{V}_{1,m} = \Big\{x_{i}\in {V}_{1}:\ \underset{y_j\in {V}_{2,m}}{\sum}\mathsf M_{i,j}\geq \underset{y_j\in {V}_{2,\ell}}{\sum}\mathsf M_{i,j}\ \textup{for any}\ \ell=1,\ldots,K \Big\}.
\end{equation}
The information extracted from the first and second sensors iteratively interacts through (\ref{equ:multiview4}) and (\ref{equ:multiview5}). This approach is called the ``co-clustering'' algorithm.
%{\color{red}Next, we summarize a generalization of the co-clustering algorithm to the multiple sensors setup, multiview DM \cite{Linderbaum_Yeredor_Salhov_Averbuch:2015}.}
%To understand the relationship of this co-clustering algorithm and the multiview DM,
We now summarize how the co-clustering of the vertices of a bipartite graph $G=({V},{E})$ is related to the traditional spectral clustering algorithm and the well known Cheeger's inequality \cite{Fan:1996}.\footnote{Its multiway clustering is supported by the recently developed theory for the multi-way spectral clustering algorithm \cite{LeeTrevisan2014}.}
Take the bi-clustering for the illustration purpose.
Take two disjoint sets $\mathcal{U}_{1}$ and $\mathcal{U}_{2}$ with
${V}=\mathcal{U}_{1}\cup\mathcal{U}_{2}$ as a set of clusters for vertices in ${V}$.
First of all, define a loss function %, {\em normal cut}, $\textup{cut}(\mathcal{U}_{1},\mathcal{U}_{2})$
%for a set of clusters $\mathcal{U}_{1},\mathcal{U}_{2}$ for vertices in ${V}$
by
\begin{equation}
\mathcal{N}(\mathcal{U}_{1},\mathcal{U}_{2}):= \frac{\textup{cut}(\mathcal{U}_{1},\mathcal{U}_{2})}{\textup{weight}(\mathcal{U}_1)}+\frac{\textup{cut}(\mathcal{U}_{1},\mathcal{U}_{2})}{\mathrm{weight}(\mathcal{U}_2)},
\end{equation}
where
\begin{equation}
\textup{cut}(\mathcal{U}_1,\mathcal{U}_2):=\sum_{i\in \mathcal{U}_{1},j\in \mathcal{U}_{2}}\mathsf M_{i,j},\quad
\textup{weight}(\mathcal{U}):=\sum_{i\in \mathcal{U}}\mathsf{D}_{i,i},
\end{equation}
and $\mathsf{D}$ is a diagonal matrix so that $\mathsf{D}_{i,i}=\sum_{j=1}^{2n}\mathsf M_{i,j}$. $\mathcal{N}(\mathcal{U}_{1},\mathcal{U}_{2})$ is usually called a {\em normalized cut}. Note that $\mathsf{D}_{i,i}$ represents the degree of the vertex $x_i\in V$ (respectively $y_{i-n}\in V$) when $i\leq n$ (respectively $i>n$).
%The weight for a set of vertices $\mathcal{U}\subset \mathcal{V}$ is defined by
The generalized partition vector $\mathbf{q}=[q_{j}]_{j=1,\ldots,2n}\in \mathbb{R}^{2n\times 1}$
is defined by
\begin{equation}
q_{j} = \left\{\begin{array}{ll}+\sqrt{\frac{\textup{weight}(\mathcal{U}_{2})}{\textup{weight}(\mathcal{U}_{1})}}, & j\in \mathcal{U}_{1},\\
-\sqrt{\frac{\textup{weight}(\mathcal{U}_{1})}{\textup{weight}(\mathcal{U}_{2})}}, & j\in \mathcal{U}_{2}.
 \end{array}\right.
\end{equation}
It is shown in \cite[Theorem 3]{dhillon2001co} that
\begin{equation}
\frac{\mathbf{q}^{\top}(\mathsf{D}-\mathsf M)\mathbf{q}}{\mathbf{q}^{\top}\mathsf{D}\mathbf{q}}
=\frac{\textup{cut}(\mathcal{U}_{1},\mathcal{U}_{2})}{\mathrm{weight}(\mathcal{U}_{1})}+\frac{\textup{cut}(\mathcal{U}_{1},\mathcal{U}_{2})}{\mathrm{weight}(\mathcal{U}_{2})}\,,
\end{equation}
which implies that the problem of finding a balanced partition with small cut value can be relaxed and cast as an eigenvalue problem as follows
\begin{equation}\label{equ:multiview6}
\underset{\mathbf{q}\neq \mathbf{0}}{\min}\frac{\mathbf{q}^{\top}(\mathsf{D}-\mathsf M)\mathbf{q}}{\mathbf{q}^{\top}\mathsf{D}\mathbf{q}},\
\textup{subject to}\ \mathbf{q}^{\top}\mathsf{D}[1\ 1\ \cdots 1]^{\top}=0.
\end{equation}
The minimizer of (\ref{equ:multiview6}) is the eigenvector {$q_2$} of $\mathsf{D}^{-1}\mathsf M$ corresponding to the second largest eigenvalue, and the bipartition is achieved by running {k-means} on $\{q_2(i)\}_{i=1}^{2n}\subset \mathbb{R}$ with $k=2$. Note that the first $n$ entries of $q_2$ (as well as other eigenvectors) are associated with the clustering of the first sensor, and the last $n$ entries are associated with the clustering of the second sensor. The result is the ``co-cluster'' of the two sensors.
%
%To co-cluster vertices associated with two sensors into more than two clusters, or usually called multiway clustering, we can directly generalize the whole idea, including the normal cut functional and the spectral clustering. Then take $2k$ eigenvectors, $q_2,\ldots,q_{2k+1}$, into account and run k-mean on $\{[q_2(i),\ldots,q_{2k+1}(i)]^\top\}_{i=1}^{2J}\subset \mathbb{R}^{2k}$. The theoretical guarantee based on the generalized Cheeger's inequality is provided by  \cite{LeeTrevisan2014}.

The co-clustering algorithm is intimately related to the recently proposed sensor fusion algorithm, multiview DM \cite{Linderbaum_Yeredor_Salhov_Averbuch:2015}, particularly when there are two sensors. This relationship is clear after we summarize the multiview DM when there are two sensors.
Form two affinity matrices
$W_{xy}:=W_{x}W_{y}\in \mathbb{R}^{n\times n}$ and
$W_{yx}:=W_{y}W_{x}\in \mathbb{R}^{n\times n}$, where $W_x$ and $W_y$ are affinity matrices for $\mathcal{X}$ and $\mathcal{Y}$ respectively that are defined as that in \eqref{Definition W matrix}.
Then, define the affinity matrix $\mathsf M$ by taking the product of affinities of two sensors by
\begin{equation}\label{equ:multiview2}
\mathsf M =
\begin{bmatrix}
0_{n\times n} & W_{xy}\\
W_{yx} & 0_{n\times n}
\end{bmatrix}\in \mathbb{R}^{2n\times 2n}\,.
\end{equation}
%where $W_x$ and $W_y$ are the affinity matrices corresponding to the first and second channels, respectively. For example, they can be the affinity matrices constructed in \eqref{Definition W matrix} for DM.
Note that the $(i,j)$-th entry of $W_xW_y$ describes how similar the information of {the} $i$-th sample captured by the $x$ sensor {is to} the information of {the} $j$-th sample captured by the $y$ sensor. Denote $q_l$ to be the $l$-th eigenvector of $\mathsf{D}^{-1}\mathsf M$.
By the above discussion, we know that $q_2(i), \cdots\, q_{\hat{d}+1}(i)$ provide the co-clustering information of the $i$-th sample captured by the first sensor, and $q_2(n+i), \cdots,q_{\hat{d}+1}(n+i)$ provide the co-clustering information of the $i$-th sample captured by the second sensor. Since both channels provide information, for each $i\in \{1,\ldots,n\}$, we consider a concatenation of both, $[q_2(i)\ \cdots\ q_{\hat{d}+1}(i)\ q_2(n+i)\ \cdots\ q_{\hat{d}+1}(n+i)]^\top$, to be the new feature of the $i$-th sample. This approach is the multiview DM algorithm proposed in \cite{Linderbaum_Yeredor_Salhov_Averbuch:2015} when there are two sensors.
Note that the multiview DM is more general than simply co-clustering since it can fuse information from multiple sensors. Its theoretical property and its relationship with other algorithms will be explored in the future work.

\section{Proposed algorithm to explore intrinsic geometry of sleep dynamics and predict sleep stage}\label{sec:proposed algorithm}

The proposed algorithm is based on the above-mentioned unsupervised feature extraction from the EEG signal. %that depends on the diffusion geometry based sensor fusion tools, DM and ADM.
The feature extraction consists of two steps. First, we extract spectral information from the EEG signal (indicated by Part 1 in Figure \ref{FlowChart}). Second, we apply DM or ADM (indicated by Part 2-1 and Part 2-2 in Figure \ref{FlowChart}) with the local Mahalanobis distance to determine the final features. We can explore the sleep dynamics by visualizing the final features.
For the sleep stage prediction purpose, we take the well established HMM to build up the prediction model.
Below, we detail the algorithm implementation step by step.

\tikzstyle{line} = [draw, -latex']
\tikzstyle{arrow} = [thick,->,>=stealth]

\begin{figure}[h!]
\centering
   \begin{tikzpicture}[>=latex']
        \tikzset{
        block/.style= {draw, rectangle, align=center,minimum width=5cm,minimum height=.10cm,line width=0.05mm},
        rblock/.style={draw, shape=rectangle,rounded corners=1.5em,align=center,minimum width=12cm,minimum height=.10cm},
        }

        \node [block]  (EEG1start) {EEG, channel 1};
        \node [block, below =.5cm of EEG1start] (EEG1Step1) {Synchrosqueezing transform:\\ {\em Synchrosqueezed EEG spectrum}};
        \node [block, below =.5cm of EEG1Step1] (EEG1Step2) {DM with the local Mahalanobis distance \\{\em Intrinsic sleep features}: \\  \includegraphics[scale=0.44]{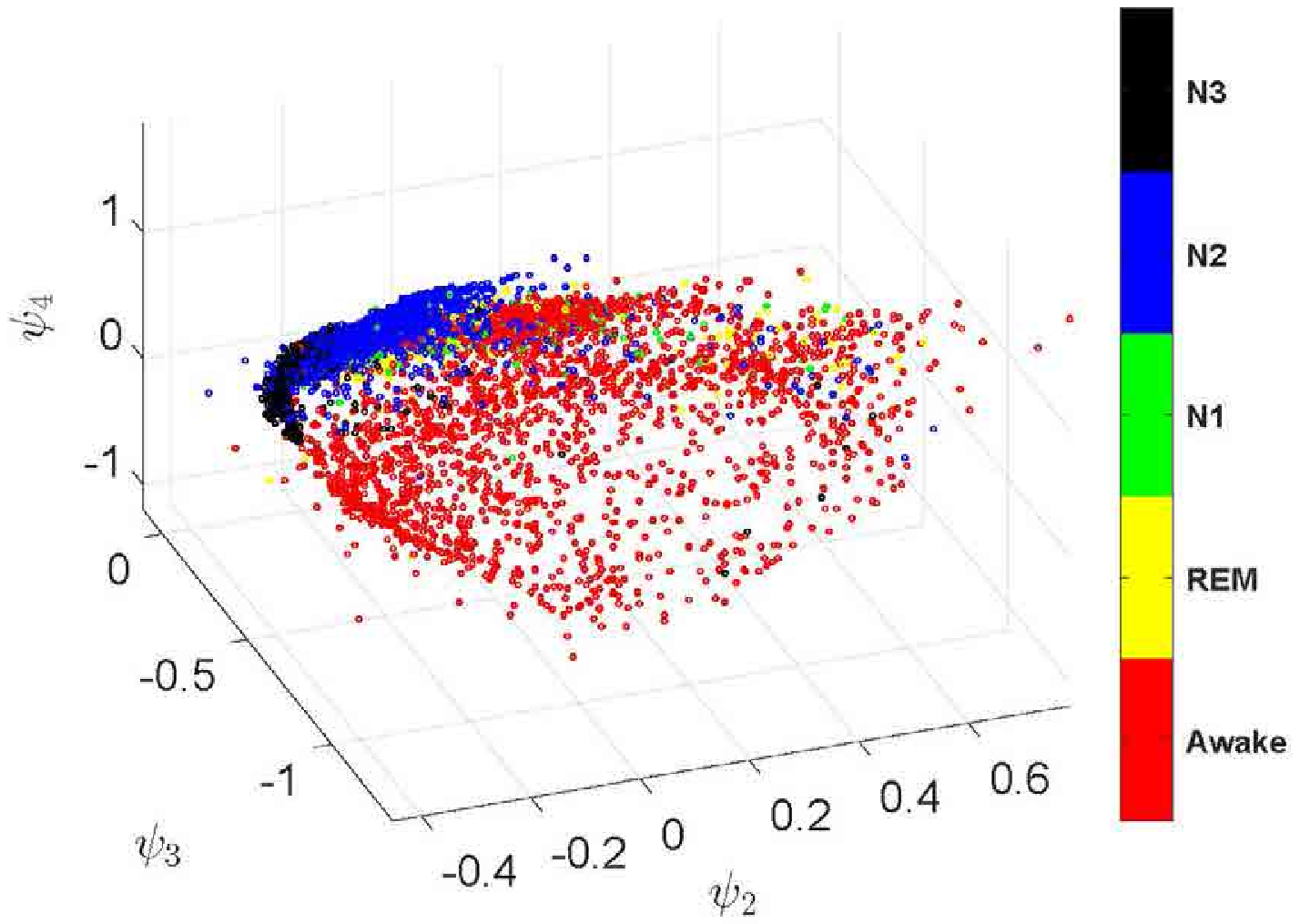}
        };

        \node [block, right =1.8cm of EEG1start]  (EEG2start) {EEG, channel 2};
        \node [block, below =.5cm of EEG2start] (EEG2Step1) {Synchrosqueezing transform:\\\em{Synchrosqueezed EEG spectrum}};
        \node [block, below =.5cm of EEG2Step1] (EEG2Step2) {DM with the local Mahalanobis distance \\ {\em Intrinsic sleep features}:  \\\includegraphics[scale=0.44]{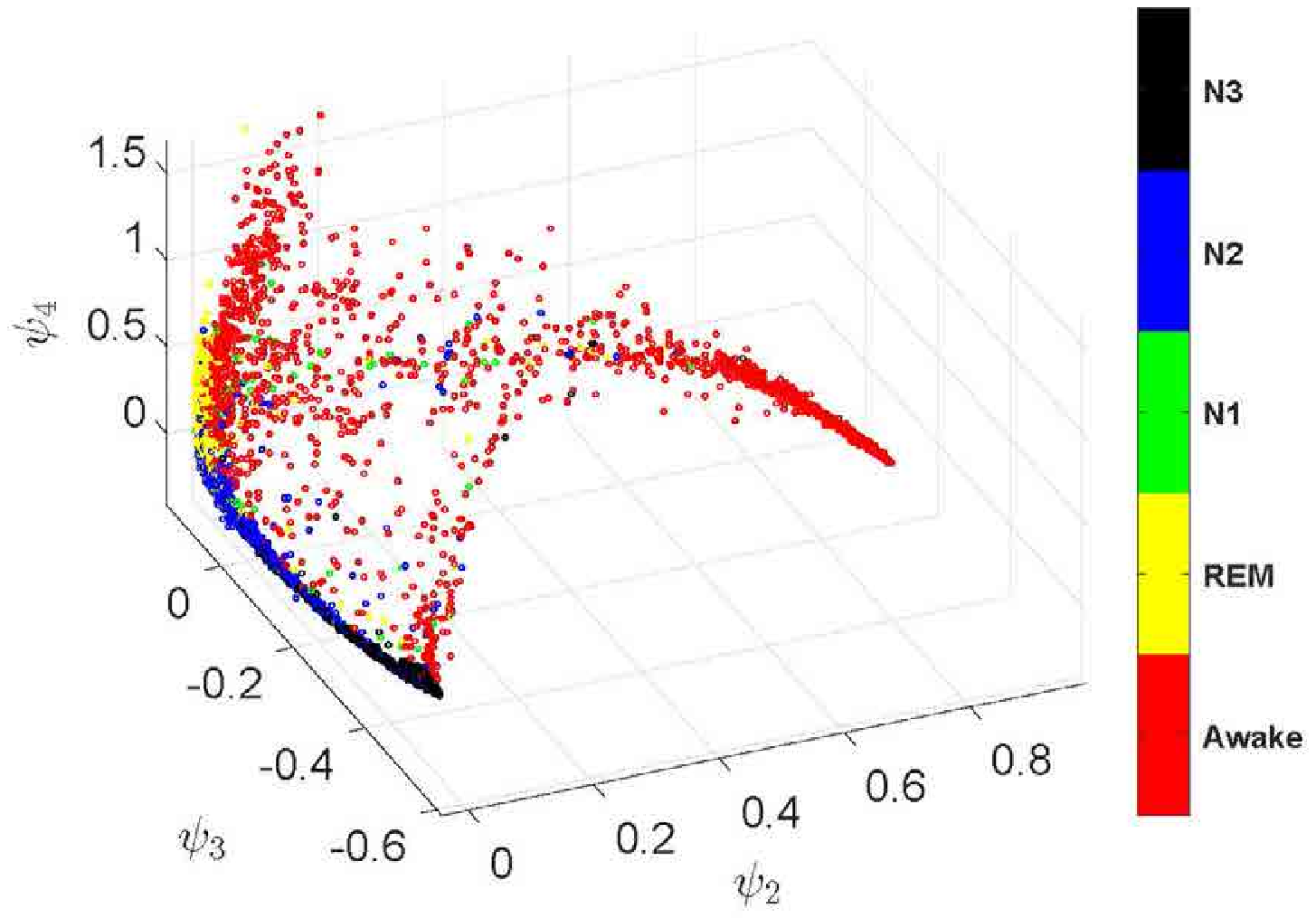}
        };

        \node [block, below right =.7cm and -3.3 cm of EEG1Step2] (Final) {sensor fusion by ADM and multiview DM \\ {\em Common intrinsic sleep features}: \\
        \includegraphics[scale=0.44]{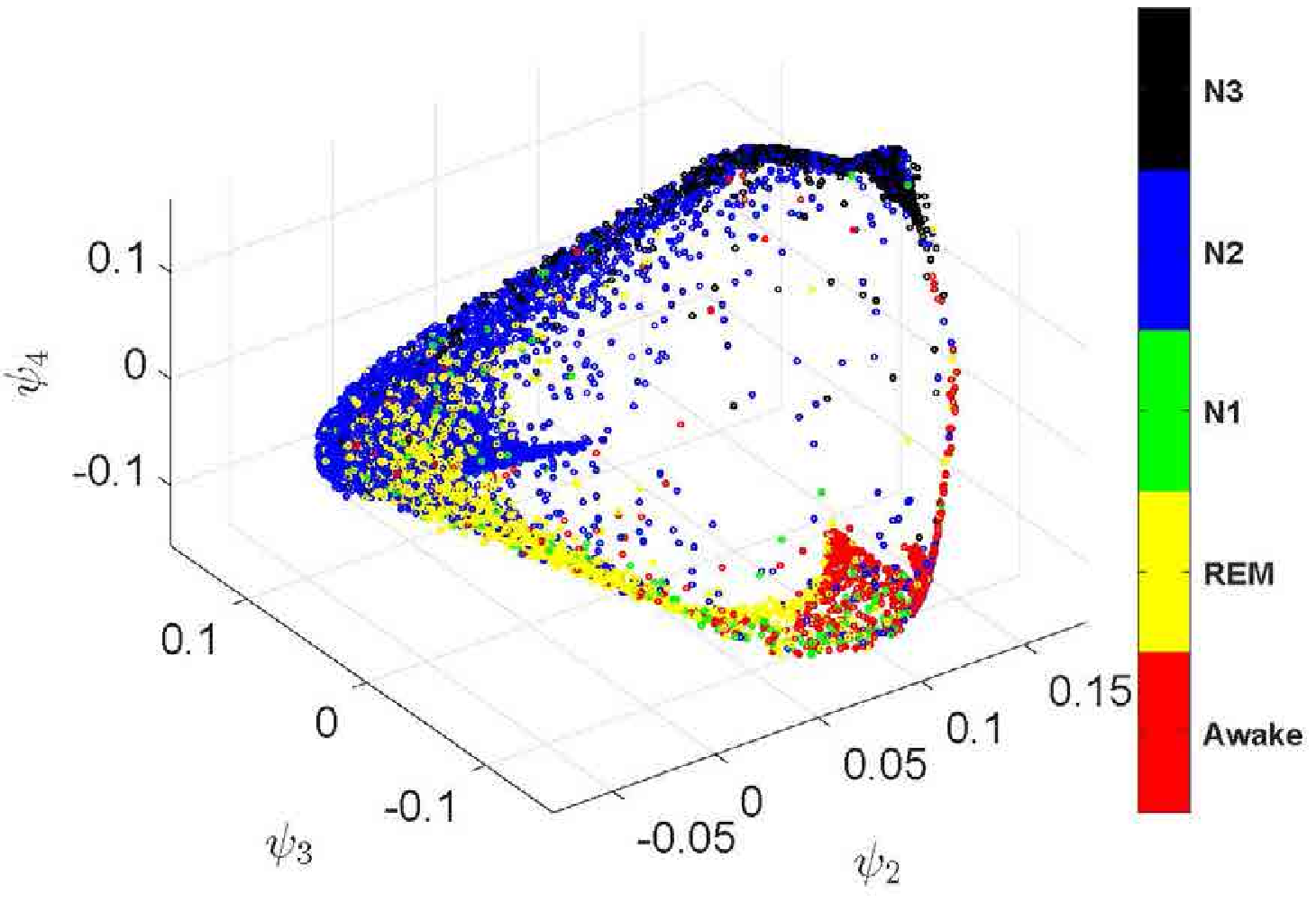}
        };

          \draw[arrow]   (EEG1start.south)  --(EEG1Step1.north) node [pos=0.66,right] {};
          \draw[arrow]   (EEG1Step1.south)  --(EEG1Step2.north) node [pos=0.66,right] {};
          \draw[arrow]   (EEG1Step2.south)  -- (Final.north) node [pos=0.66,right] {};

                    \draw[arrow]   (EEG2start.south)  --(EEG2Step1.north) node [pos=0.66,right] {};
          \draw[arrow]   (EEG2Step1.south)  --(EEG2Step2.north) node [pos=0.66,right] {};
          \draw[arrow]   (EEG2Step2.south)  -- (Final.north) node [pos=0.46,right] {};

    \end{tikzpicture}
    \caption{\label{FlowChart}The flow chart of the proposed feature extraction steps and a visualization of the extracted features by the diffusion map and sensor fusion are shown. The signal is from subject SC414 in Sleep-EDF SC$^{*}$ \cite{Goldberger_Amaral_Glass_Hausdorff_Ivanov_Mark_Mietus_Moody_Peng_Stanley:2000}, where Channel 1 is Fpz-Cz and channel 2 is Pz-Oz. In the bottom figure, only the result of AD is shown. The ratios of the stages Awake, REM, N1, N2, and N3 are 14.8\%, 21.9\%, 5.7\%, 47.5\%, and 10.1\%.}
\end{figure}

\subsection{Step 1: Extract synchrosqueezed EEG spectral feature}

Take $x\in \mathbb{R}^n$ to be the digitalized EEG signal sampled uniformly every $\tau$ second from a subject during his/her sleep,
where $\tau$ is the reciprocal of the sampling rate.
For $j=1,2,\ldots$, the STFT at {$(j\tau)$ seconds} is directly implemented by the weighted Fourier transform:
\begin{equation}\label{STFT_discrete}
\mathbf{V}^{(h)}_{x}(j,k):=\sum^{j+\lceil 5/\tau\rceil}_{m=j-\lfloor 5/\tau\rfloor } x_{m} \frac{1}{H}h\Big(\frac{m}{H}\Big)e^{-i2\pi \frac{k}{K} m},
\end{equation}
where {$\lceil x\rceil$ means the smallest integer greater than $x>0$, $\lfloor x\rfloor$ is the largest integer smaller than $x>0$,} $k\in\{0,1,\ldots,K-1\}$, $K\in \mathbb{N}$ is a parameter used to adjust the frequency resolution, $h(z)$ is the standard Gaussian function, and $H>0$ is the bandwidth. The SST is then implemented as
\begin{align}\label{synchro_discrete}
\mathbf S_x(j,\hat{k}) =\underset{k\in \Lambda(\hat{k})}{\sum} |\mathbf{V}^{(h)}_{x}(j,k)|^2
\end{align}
where
\begin{align}\notag
\Lambda(\hat{k})=
\left\{k\in \{0,1,\ldots,K-1\}\Big|\,  k - \Im\left(\frac{K}{2\pi H}\frac{\mathbf{V}^{(\mathcal{D}h)}_{x}(j,k)}{\mathbf{V}^{(h)}_{x}(j,k)}\right)\in \Big[\hat{k}-\frac{1}{2},\hat{k}+\frac{1}{2}\Big)\right\}
\end{align}
and $\mathcal{D}h$ is the derivative of $h$. Note that $\mathbf S_x(j,\cdot)\in \mathbb{R}^K$ is the synchrosqueezed EEG spectrogram at {$(j\tau)$ seconds}, which can be viewed as a dynamical spectral feature for the sleep dynamics. Note that the numerical algorithm is a direct discretization of the continuous setup for SST in Section \ref{sec:SST}, and hence the synchrosqueezed EEG spectrogram.

Finally, we {follow the common dimension reduction approach to} convert the synchrosqueezed EEG spectrogram into the features {that interest us}. In this work, we follow the sleep stage scoring standard, the AASM criteria \cite{Iber2007}, and take an epoch for the sleep stage evaluation to be 30 seconds long.
Therefore, we have {$J:=n\tau/30$} epochs.
{Consider 9 frequency regions $R_1,\ldots, R_9\subset\mathbb{R}$ in the spectral domain, defined as $R_{1}=[0.5,4]$ ($R_1$ is the delta band), $R_{2}=[4,7]$ ($R_2$ is the theta band), $R_{3}=[7,12]$ ($R_3$ is the alpha band), $R_{4}=[12,16]$ ($R_4$ catches the spindle), $R_{5}=[16,20]$,
$R_{6}=[20,24]$, $R_{7}=[24,28]$, $R_{8}=[28,31]$ ($R_5$ to $R_8$ form the beta band), and $R_{9}=[31,49]$ ($R_9$ is the gamma band). These bands are chosen due to their well known physiological meanings \cite{campbell2009eeg}. In practice we observed that suitably dividing the beta wave frequency range and the low-gamma wave frequency range ($25\sim 49$ Hz) into finer bands leads to higher classification accuracy, so we consider the frequency bands $\{R_{\ell}\}_{\ell=1,\ldots,9}$ in this work.}

For the $j$-th epoch, we get a 10-dimension vector ${\bf u}^{(j)}=[u_{0}^{(j)}\,\, u_{1}^{(j)}\,\, \cdots\, u_{9}^{(j)}]$ that include the total energy
\begin{equation}\label{eq3}
u_{0}^{(j)}= \frac{\tau}{30} \underset{\hat{k}:\hat{k}/(\tau K)\in [0.5,49]}{\sum}\ \overset{30j/\tau}{\underset{\hat{j}=30(j-1)/\tau+1}{\sum}}{\bf S}_{x}\big(\hat{j},\hat{k}\big)
\end{equation}
and the {band power} ratios on $R_1,\ldots, R_9$:
\begin{equation}\label{eq4}
u^{(j)}_{\ell} = \frac{\tau}{30} \frac{1}{u_{0}^{(j)}}\underset{\hat{k}:\hat{k}/(\tau K)\in R_{\ell}}{\sum}\ \overset{30j/\tau}{\underset{\hat{j}=30(j-1)/\tau+1}{\sum}}{\bf S}_{x}\big(\hat{j},\hat{k}\big),\ \ell=1,\ldots,9.
\end{equation}
We call $\mathbf{u}^{(j)}$, $j=1,\ldots,J$, the {\em synchrosqueezed EEG spectral feature} of the $j$-th epoch.

\subsection{Step 2: convert synchrosqueezed EEG spectral feature into intrinsic sleep feature}

In the optimal situation, the {spectral content} of the sleep dynamics can be {well captured} by the synchrosqueezed EEG spectral features. However, the synchrosqueezed EEG spectral features might be erroneous due to the inevitable noise, other sensor-specific artifacts and the information distortion caused by the observation procedure. We then stabilize these features to better quantify the intrinsic sleep dynamics

Take the synchrosqueezed EEG spectral features $\mathcal{U}^x:=\{\mathbf{u}^{(j)}\}_{j=1}^{J}$, which is a point cloud in an Euclidean space.
First, from the point cloud $\mathcal{U}^x$, we build a graph with $\mathcal{U}^x$ being vertices. The affinity between the features $\mathbf{u}^{(i)}$ and $\mathbf{u}^{(j)}$ is defined
\begin{equation}\label{affinity_matrix}
W_x(i,j)=\exp\left\{-\frac{d_{\texttt{LMD}}^{2}(\mathbf{u}^{(i)},\mathbf{u}^{(j)})}{\varepsilon}\right\},\ \ \textup{for}\ \ i,j=1,\ldots,J,\ i\neq j,
\end{equation}
where $\epsilon>0$ is chosen by the user and $d_{\texttt{LMD}}(\cdot,\cdot)$ is the local MD.
The local MD is chosen due to its scale-invariant property and stability property \cite{TalmonPNAS}.
To calculate the local MD,
denote the $K$-neighborhood of $\mathbf{u}^{(j)}$ by $\mathcal{N}_j$ for each $j\in\{1,2,\ldots,J\}$, where $K=[\alpha J]$ and the ratio $\alpha$ is predetermined and calculate the local covariance matrix $\Gamma_j$ defined in \eqref{cov2}. Then, evaluate the local MD by \eqref{Definition:EIG_dis2}.
With the $J\times J$ affinity matrix $W_x$, the degree matrix $D_x$ of size $J\times J$ is constructed as that in \eqref{degree_matrix}, and hence
the DM $\Phi^x_t$, where $t>0$.
As a result, the synchrosqueezed EEG spectral features $\mathcal{U}^x$ are converted into a set of new features $\Phi^x_t(\mathbf{u}^{(j)})\in \mathbb{R}^{\hat{d}}$.
We call them the {\em intrinsic sleep features} for the EEG signal $x$, and
denoted
\begin{equation}\label{Definition:intrinsic sleep features}
\mathcal{F}^x:=\{\Phi^x_t(\mathbf{u}^{(j)}) \}_{j=1}^J\subset \mathbb{R}^{\hat{d}}.
\end{equation}
See Figure \ref{FlowChart} for an example of DM when $\hat{d}=3$. It is clear that epochs of different sleep stages are clustered and separated, and this shows the reason we call $\mathcal{F}^x$ the intrinsic sleep features.

\subsection{Step 3: fuse two intrinsic sleep features to common intrinsic sleep feature}

When we have two simultaneously recorded EEG channels $x$ and $y$, for each channel, we obtain its intrinsic sleep features, denoted as $\mathcal{F}^x \subset \mathbb{R}^{\hat{d}_x}$ and $\mathcal{F}^y \subset \mathbb{R}^{\hat{d}_y}$ respectively, where $\hat{d}_x$ might be different from $\hat{d}_y$. Via AD, $A=A_{x}A_{y}$, we obtain the common metric between channels $x$ and $y$. Then, run DM (\ref{affinity_matrix}) with the common metric between the $i$-th epoch and the $j$-th epoch. Denote the first $\hat{d}$ nontrivial eigenvectors of the associated transition matrix by $\psi_2,\ldots,\psi_{\hat{d}+1}$, where $\hat{d}$ is chosen to be $10$ in this work.
For the co-clustering, we choose
\begin{equation}\label{equ:multiview2}
\mathsf M =
\begin{bmatrix}
0_{J\times J} & W_xW_y\\
W_yW_x & 0_{J\times J}
\end{bmatrix}\in \mathbb{R}^{2J\times 2J},
\end{equation}
{where $W_x$ and $W_y$ are affinity matrices defined in \eqref{affinity_matrix}}, and denote a diagonal matrix $\mathsf{D} \in \mathbb{R}^{2J\times 2J}$ so that its $i$-th diagonal entry is the sum of the $i$-th row of $\mathsf M$. Denote $q_i\in \mathbb{R}^J$ to be the $i$-th left eigenvector of the transition matrix $\mathsf{D}^{-1}\mathsf M$.
Since for each $i$, $q_i(l)$ and $q_i(J+l)$ correspond to the $l$-th epoch for each $l\in\{1,\ldots,J\}$, we could consider the $2\tilde{d}$ vector, $[q_2(j)\ \cdots\ q_{\tilde{d}+1}(j)\ q_2(J+j)\ \cdots\ q_{\tilde{d}+1}(J+j)]$, to be another set of features associated with the sleep stage of the $j$-th epoch, where $j\in \{1,\ldots,J\}$.
With AD and co-clustering, call the $\hat{d}+2\tilde{d}$ dimensional vector
\begin{align}
v_{j}:=[\psi_2(j),\ldots,\psi_{\hat{d}+1}(j) ,q_2(j),\cdots, q_{\tilde{d}+1}(j), q_2(J+j), \cdots, q_{\tilde{d}+1}(J+j)]^\top\label{Definition:common intrinsic sleep feature}
\end{align}
the {\em common intrinsic sleep feature} associated with the $j$-th epoch. Denote
\[
\mathcal{F}^{x,y}:=\{v_j\}_{j=1}^J\subset \mathbb{R}^{\hat{d}+2\tilde{d}}.
\]
An illustration of the result of AD with $\hat{d}=3$ is shown in Figure \ref{FlowChart}.

\subsection{Learning step: Sleep Stage Classification by the Hidden Markov Model}\label{sec:HMM}

To predict sleep stage, we choose the standard and widely used algorithm HMM for the classification purpose. HMM is particularly powerful if we want to model a sequence of variables that changes randomly in time.
Although it is standard, to make the paper self-contained, we provide a summary of the HMM and its numerical implementation in this section.

In general, a HMM can be viewed as a doubly embedded stochastic sequence with a sequence that is not observable (hidden-state sequence) and can only be observed through another stochastic sequence (observable sequence). An HMM can be fully specified by the hidden-state space, the hidden-state transition matrix, the observation space, the emission probability matrix, and the initial status. From the training dataset, the HMM could be established as a prediction model for the testing dataset. Below, we provide a summary of the HMM and its numerical details.

The {\em hidden-state space} $\mathcal{S}$ consists of five sleep stages: Awake, REM, N1,  N2, and N3. To simply the notation, we label Awake, REM, N1, N2 and N3 by 1, 2, 3, 4 and 5, respectively; that is,  $\mathcal{S}=\{1,\ldots,5\}$. The sleep stage on the $j$-th epoch is viewed as a random variable $S_{j}$, whose realization is denoted by $s_{j}\in \mathcal{S}$.
{To collect high-quality and reliable EEG signals,
the calibration is generally carried out. During the calibration period,
the testing subject is awake. Hence, the recording starts from an Awake epoch, which implies that
$S_{0}=1$. }
Assume that the time series $\{S_{0},S_{1},S_{2},...\}$ is homogeneous and the hidden-state transition matrix $\mathfrak{M}:=(m_{ij})_{1\leq i,j\leq 5}$ satisfies
\begin{equation}\label{matrix_aij}
m_{ij}=P(S_{t+1}=j\mid S_{t}=i).
\end{equation}
By the homogeneous assumption, the probability of hidden-state transition on the right hand side of (\ref{matrix_aij}) can be estimated by the number of transitions from state $i$ to state $j$ normalized by the number of transitions from state $i$, i.e., $\hat{m}_{ij}= \frac{\#\{t:\,s_{t}=i,s_{t+1}=j\}}{\#\{t:\,s_{t}=i\}}$.

The common intrinsic sleep features $\mathcal{F}^{x,y}=\{v_{j}\}_{j=1}^{J}$ can be viewed as our observation for the hidden-state sequence $\{s_{j}\}_{j=1}^{J}$. We create a codebook to quantize $\{v_{j}\}_{j=1}^{J}$ and define the {\em observation state space} by the Linde-Buzo-Gray (LBG) algorithm \cite{buzo1980speech}.
Denote the codebook as $\mathcal O$ for the observation state space, which is represented by symbols $\{1,2,\ldots,|\mathcal O|\}$, where $|\mathcal O|$ is the cardinality of $\mathcal O$. Based on the above vector quantization, we have an observable time series $\{O_1,O_2,\ldots,O_J\}$, where $O_{j}$ is the random variable describing the observation at the $j$-th epoch, which takes a value from the codebook $\mathcal O$. The realization of $O_j$ is denoted as $o_j$.

Consider the emission probability $\mathsf{E}:=(e_{j}(k))_{j\in\mathcal{S}, k\in\mathcal{O}}$ to quantify the probability of observing state $k$ from the $j$-th hidden state, and assume the following relationship
\begin{equation}\label{matrix_b}
e_{j}(k)=P(O_{t}=k\mid S_{t}=j).
\end{equation}
Based on the time-homogeneous assumption, $\mathsf{E}$ can be estimated by the accumulated number of times in hidden state $j$ and simultaneously observing symbol $k$ normalized by the number of times in hidden state $j$; that is, $\hat{e}_{j}(k)= \frac{\#\{t:\,s_{t}=j,o_{t}=k\}}{\#\{t:\,s_{t}=j\}}$.
With the initial distribution $\pi_{0}, \{  m_{ij}\} $ and $\{ e_{j}(k)\} $, we have
\begin{align}
&P(S_{0}=s_{0},S_{1}=s_{1},\ldots,S_{J}=s_{J},O_{1}=o_{1},\ldots,O_{J}=o_{J})\nonumber\\
=&\,\pi_{0}(s_{0})m_{s_{0}s_{1}}e_{s_{1}}(o_{1})
  \ldots m_{s_{J-1}s_{J}}e_{s_{J}}(o_{J}).
\end{align}
Also we have the Markov property  for this chain.

Given the trained HMM, that is, the estimated hidden-state transition matrix, the estimated emission probability matrix, and the initial status,  we now detail an algorithm to estimate the sleep stages of the testing subject $\mathcal T:=\{v_1,\ldots,v_J\}$, where $J$ is the number of epochs.
By the codebook, $\mathcal T$ is vector-quantized into $\{o_{1},\ldots,o_J\}$. By Markov property, our goal is to find the most possible sequence of hidden states for the testing subject
i.e., a path $\left(s^{*}_{1},\ldots,s^{*}_{J}\right)$ that maximizes the probability
\begin{equation}
P(S_{1}=s_{1},\ldots,S_{J}=s_{J},O_{1}=o_{1},\ldots,O_{J}=o_{J})\,.
\end{equation}
Here we assume that $s_{0}=1$ (awake) and $P(O_{J+1}=o_{J+1},S_{J+1}=F)=1$ for some  $F$ in the hidden state space and some $o_{J+1}$ in the observation state space.

Define
\begin{equation}\label{def:v}
\nu_t(j) =
\underset{s_{1},\ldots,s_{t-1}}{\max}P(s_{1},\ldots,s_{t-1},o_{1},\ldots,o_{t},S_{t}=j)
\end{equation}
for $t=1,2,\ldots,J+1$, which represents the maximal probability that the HMM is in state $j$ at time $t$ and $o_{1},\ldots,o_{t}$ are the observations up to time $t$.
The maximum in (\ref{def:v}) is taken over all probable state sequence $s_{1},\ldots,s_{t-1}$.
Note that when $t=J+1$,
\begin{align}
\nu_{J+1}(F) =&
\underset{s_{1},\ldots,s_{J}}{\max}P(s_{1},\ldots,s_{J},o_{1},\ldots,o_{J+1},S_{J+1}=F)
\nonumber\\
=& \underset{s_{1},\ldots,s_{J}}{\max}P(s_{1},\ldots,s_{J},o_{1},\ldots,o_{J+1})
=
P(s^{*}_{1},\ldots,s^{*}_{J},o_{1},\ldots,o_{J+1}),\label{v_eq}
\end{align}
where the last equality holds {for some} $s^{*}_{1},\ldots,s^{*}_{J}\in \mathcal{S}$.
This fact motivates us the following algorithm to find $\nu_t(j)$.
First of all, for $t=1$, $\nu_1(j)=P(o_{1},S_{1}=j)=m_{s_{0}j}e_{j}(o_{1})=m_{1j}e_{j}(o_{1})$, where $j\in \mathcal{S}$.
For $t=2$,
\begin{align}
\nu_{2}(j) &=
\underset{s_{1}}{\max}P(s_{1},o_{1},o_{2},S_{2}=j)\nonumber\\
&= \underset{s_{1}}{\max} \ m_{1s_{1}}e_{s_{1}}(o_{1})m_{s_{1}j}e_{j}(o_{2})
\label{v2}= \underset{s_{1}}{\max} \ \nu_{1}(s_{1})m_{s_{1}j}e_{j}(o_{2}).
\end{align}
Denote the maximizer of the right hand side of (\ref{v2}) by $V(j,2)$, that is,
\[
V(j,2)= \arg\underset{s_1}{\max} \ P(S_{1}=s_1,o_{1},o_{2},S_{2}=j)\in \mathcal{S}.
\]
Note that since $V(j,2)=\arg\underset{s_1}{\max} \ P(S_{1}=s_1\mid o_{1},o_{2},S_{2}=j)$, $V(j,2)$ is the most likely state at time index $1$ when the state is in $j$ at time index $2$ {and} when $o_{1}$ and $o_{2}$ are observed.
In general, we have
\begin{align}\label{v4}
\nu_{t+1}(j)=
 \underset{s_{t}}{\max} \ \nu_{t}(s_{t})m_{s_{t}j}e_{j}(o_{t+1})
\end{align}
for $t=2,\ldots,J$ and $j\in \mathcal{S}$.
Denote the maximizer of the right hand side of (\ref{v4}) by $V(j,t+1)$, which can be interpreted as the best {\em relay point} connecting the node $S_{t+1}=j$ with the most likely path that emits the symbols $o_{1},\ldots,o_{t+1}$.
With $\{V(\cdot,t)\}_{t=1,\ldots,J+1}$, we can find the optimal path
from the state $F$ at time index $J+1$ iteratively as follows:
\begin{align*}
s^{*}_{J}&=V(F,J+1)\,,\quad
s^{*}_{J-1}=V(s^{*}_{J},J)\ldots s^{*}_{2}=V(s^{*}_{3},3)\,,\quad
s^{*}_{1}=V(s^{*}_{2},2)\,.
\end{align*}
For details, see \cite{fraser2008hidden}.

\section{Material and Statistics}\label{sec:description_database}

To evaluate the proposed algorithm, we consider a publicly available database and follow standard performance evaluation procedures.

\subsection{Material}
To evaluate the proposed algorithm, we consider the commonly considered benchmark database, Sleep-EDF Database [Expanded], from the public repository Physionet \cite{Goldberger_Amaral_Glass_Hausdorff_Ivanov_Mark_Mietus_Moody_Peng_Stanley:2000}. It contains two subsets (marked as SC$^{*}$ and ST$^{*}$). The first subset SC$^{*}$ comes from healthy subjects without any sleep-related medication. The subset SC$^{*}$ contains Fpz-Cz/Pz-Oz EEG signals recorded from 10 males and 10 females without any sleep-related medication, and the age range is 25-34 year-old. There are two approximately 20-hour recordings per subject, apart from a single subject for whom there is only a single recording.
The EEG signals were recorded during two subsequent day-night periods at the subjects' home. The sampling rate is 100 Hz.
The second subset ST$^{*}$ was obtained in a 1994 study of temazepam effects on the sleep of subjects with mild difficulty falling asleep. The subset ST$^{*}$ contains Fpz-Cz/Pz-Oz EEG signals recorded from 7 males and 15 females, who had mild difficulty falling asleep. Since this data set is originally used for studying the effects of temazepam, the EEG signals were recorded in the hospital for two nights, one of which was after temazepam intake. Only their placebo nights can be downloaded from \cite{Goldberger_Amaral_Glass_Hausdorff_Ivanov_Mark_Mietus_Moody_Peng_Stanley:2000}. The sampling rate is 100 Hz.
For both SC$^*$ and ST$^*$ sets, each 30s epoch of EEG data has been annotated into the classes Awake, REM, N1, N2, N3 and N4. The epochs corresponding to movement and unknown stages were excluded and the the epochs labeled by N4 are relabeled to N3 according to the AASM standard \cite{Iber2007}.
For more details of the database, we refer the reader to \url{https://www.physionet.org/physiobank/database/sleep-edfx/}.

\subsection{Statistics}

To evaluate the performance of the automatic sleep stage annotation, we shall distinguish two common cross validation (CV) schemes. According to whether the training data and the testing data come from different subjects, the literature is divided into two groups, {\em leave-one-subject-out} and {\em non-leave-one-subject-out} CV. When the validation set and training set are determined {\em on the subject level}, that is, the training set and the validation set contain different subjects, we call it the leave-one-subject-out CV (LOSOCV) scheme; otherwise we call it the non-LOSOCV scheme. The main challenge of the LOSOCV scheme comes from the inter-individual variability, but this scheme is close to the real-world scenario -- how to predict the sleep dynamics of a new arrival subject from a given annotated database. On the other hand, in the non-LOSOCV scheme, the training set and the testing set are dependent, and the performance might be over-estimated.
To better evaluate the performance of the proposed automatic sleep scoring algorithm, we choose the {LOSOCV scheme}. For each database, one subject is randomly chosen as the testing set and the other subjects form the training set.
For the testing subject, we take the phenotype information to find the $\hat{K}$ most similar subjects to establish the HMM model.
The impact of age on the sleep dynamics \cite{Vitiello2004} and EEG signal \cite{BoselliParrino1998,VanCauter2000} is well-known, so the EEG information from subjects with similar age will provide more information.
While the sleep dynamics is influenced by other phenotype information, since age is the common information among databases we consider, we determine the {$\hat{K}$} most similar subjects by the age. Note that this approach {imitates} the real scenario -- for a new-arriving subject, we can score its sleep stages by taking the existing database with annotation into account. {Also note that this LOSOCV scheme helps prevent overfitting and fully takes the inter-individual variability into account in constructing the prediction model.}

All performance measurements used in this paper are computed through
the unnormalized confusion matrix ${ M}\in \mathbb{R}^{5\times 5}$.
For $1\leq p,q\leq 5$, the entry $M_{pq}$ represents the number of expert-assigned $p$-class epochs, which were predicted to the $q$-class.
The precision ($\textup{PR}_p$), recall ($\textup{RE}_p$), and F1-score ($\textup{F1}_p$) of the $p$-th class, where $p=1,\ldots,5$, are computed respectively through
\begin{equation}
\textup{PR}_{p}= \frac{M_{pp}}{\sum_{q=1}^5M_{qp}},\quad
\textup{RE}_{p}= \frac{M_{pp}}{\sum_{q=1}^5M_{pq}},\quad
\textup{F1}_{p}=\frac{2 \textup{PR}_{p}\cdot\textup{RE}_{p}}{\textup{PR}_{p}+\textup{RE}_{p}}\,.
\end{equation}
The overall accuracy (ACC), macro F1 score ($\textup{Macro-F1}$) and kappa ($\kappa$) coefficient are computed respectively through
\begin{equation}
\textup{ACC}= \frac{\sum_{p=1}^5  M_{pp}}{\sum_{p,q=1}^5M_{pq}   },\quad
\textup{Macro-F1}=\frac{1}{5}\sum_{p=1}^5\textup{F1}_{p},\quad
\kappa = \frac{\textup{ACC}-\textup{EA}}{1-\textup{EA}},
\end{equation}
where {EA means the expected accuracy,} which is defined by
\begin{equation}
\textup{EA}= \frac{\sum_{p=1}^5\left(   \sum_{q=1}^5 M_{pq}\right)\times \left(\sum_{q=1}^5M_{qp}\right)}{\left(\sum_{p,q=1}^5     M_{pq}\right)^{2}}.
\end{equation}

{To evaluate if two matched samples have the same mean, we apply the one-tail Wilcoxon signed-rank test under the null hypothesis that the difference between the pairs follows a symmetric distribution around zero.
When we compare the variance, we apply the one-tail F test under the null hypothesis that there is no difference between the variances.
We consider the significance level of $0.05$. To handle the multiple comparison issue, we consider the Bonferroni correction. }

\section{Results}\label{sec:experiments}

We report the results of applying the proposed algorithm to the above-mentioned two databases. {The parameters in the numerical implementation are listed here. $\tau=1/100$. For SST, we choose $H=1001$ and $K=4004$ in (\ref{STFT_discrete}) and (\ref{synchro_discrete}); that is, the bandwidth is $1001/100=10.01$ seconds and we oversample the frequency domain by a factor of $4$. For the local MD, we take $\alpha=0.1$ in (\ref{affinity_matrix}) and $d=7$ in \eqref{Definition:EIG_dis2}. For DM, the $\epsilon$ in (\ref{affinity_matrix}) is chosen to be the $5$\% percentile of pairwise distances, the diffusion time is $t=1$ and we choose $\hat{d}=10$. For the co-clustering,  $\tilde{d}$ is chosen to be $10$. No systematic parameter optimization is performed to avoid overfitting. For the reproducibility purpose, the Matlab code will be provided via request.}

\subsection{Sleep dynamics visualization}
\begin{figure}[!htb]
\centering
%\begin{minipage}{0.68\textwidth}
\subfigure
{\hspace{-0.95cm}{\includegraphics[scale=0.48]{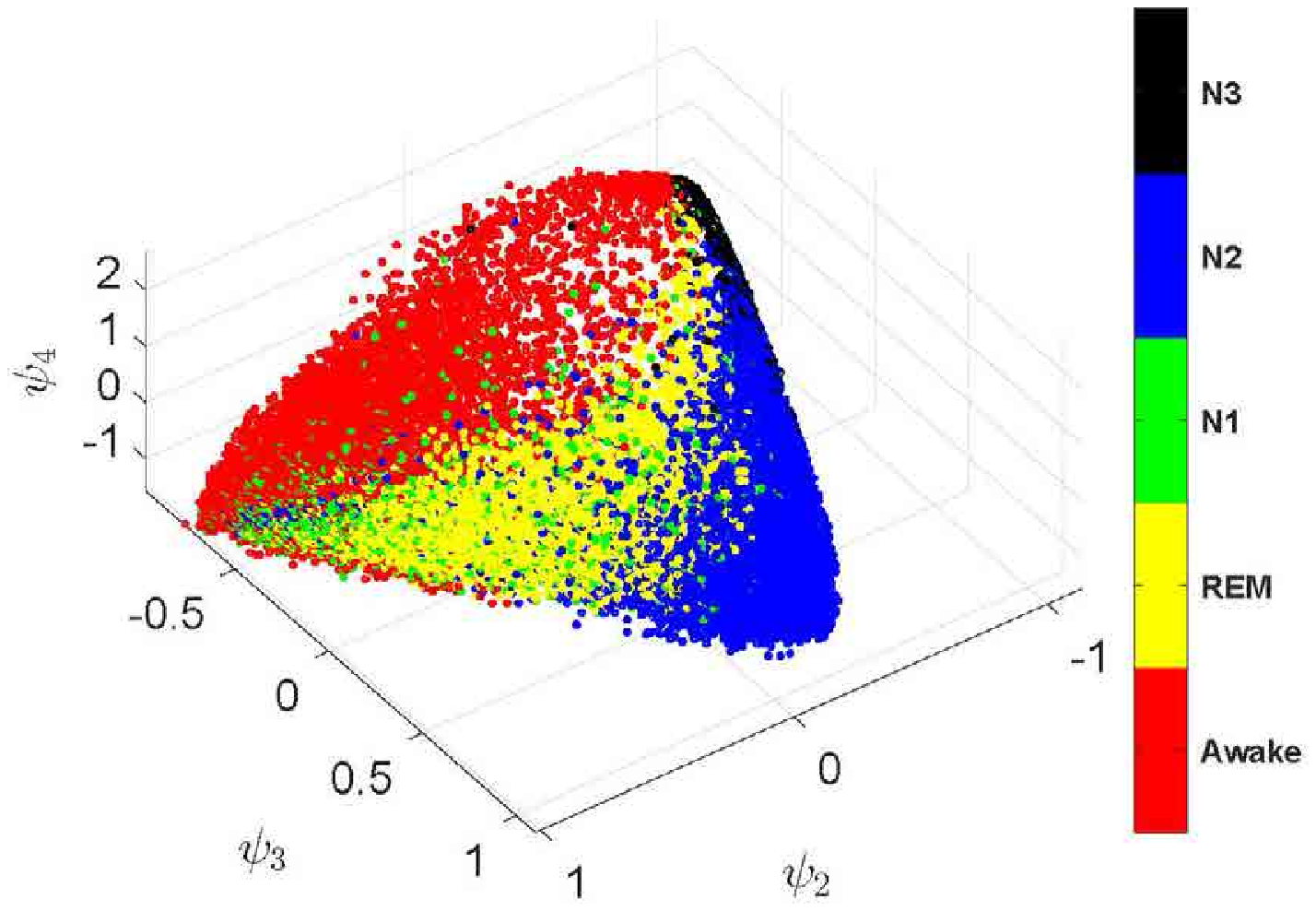}}}
\subfigure
{\hspace{-0.95cm}{\includegraphics[scale=0.48]{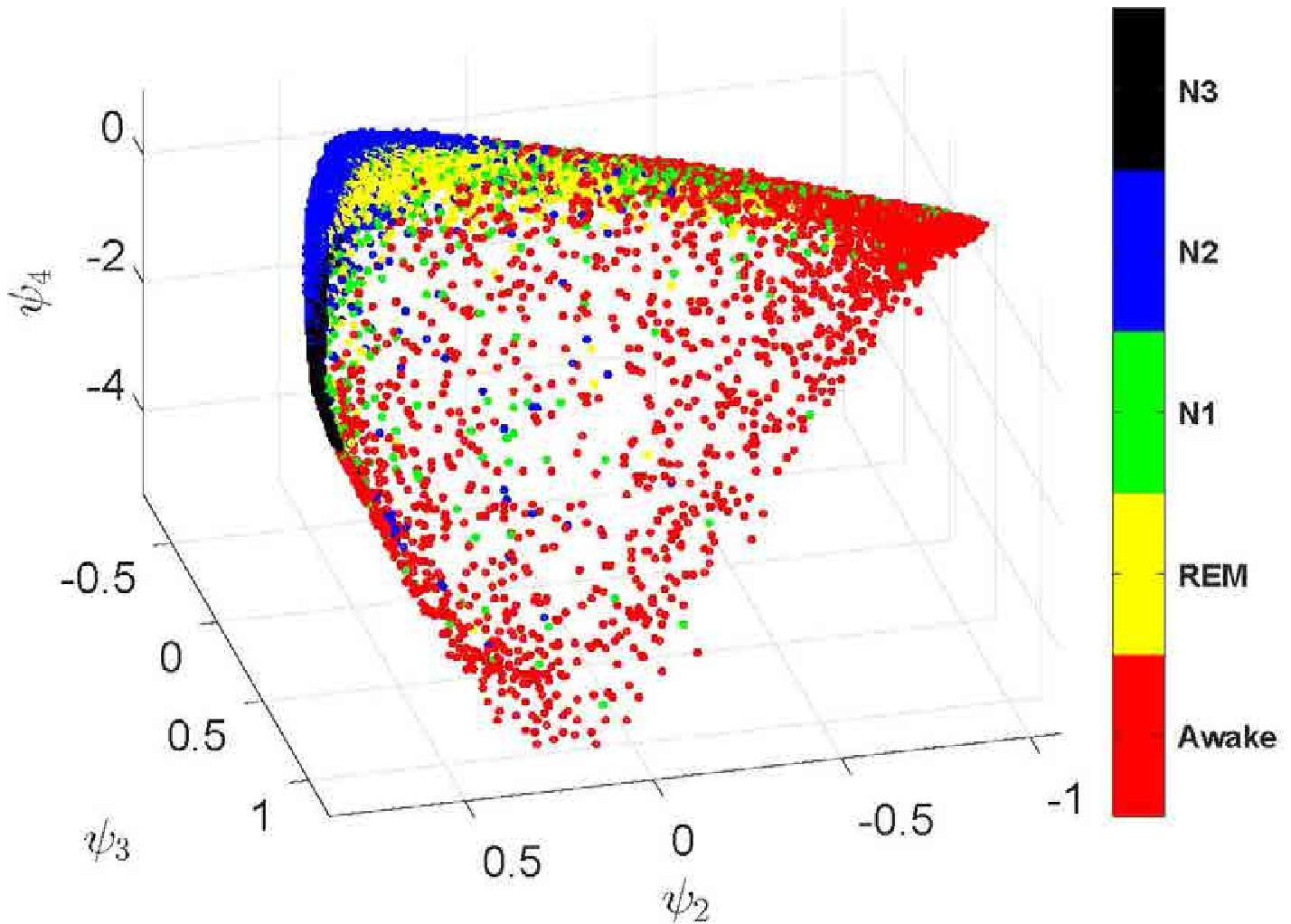}}}
%\end{minipage}
%\begin{minipage}{0.3\textwidth}
\caption{A visualization of the intrinsic sleep features (from single channel) extracted from 12 different subjects from the Sleep-EDF database (SC*).
The ratios of the stages Awake, REM, N1, N2, and N3 are
17.3\%, 18.5\%, 4.3\%, 46.0\%, and 13.9\% respectively. Each point corresponds to a 30-second epoch.\label{fig:SCvisulization0qq}
}
%\end{minipage}
\end{figure}

\begin{figure}[!htb]
\centering
%\begin{minipage}{0.68\textwidth}
\subfigure
{\hspace{-0.95cm}\includegraphics[scale=0.48]{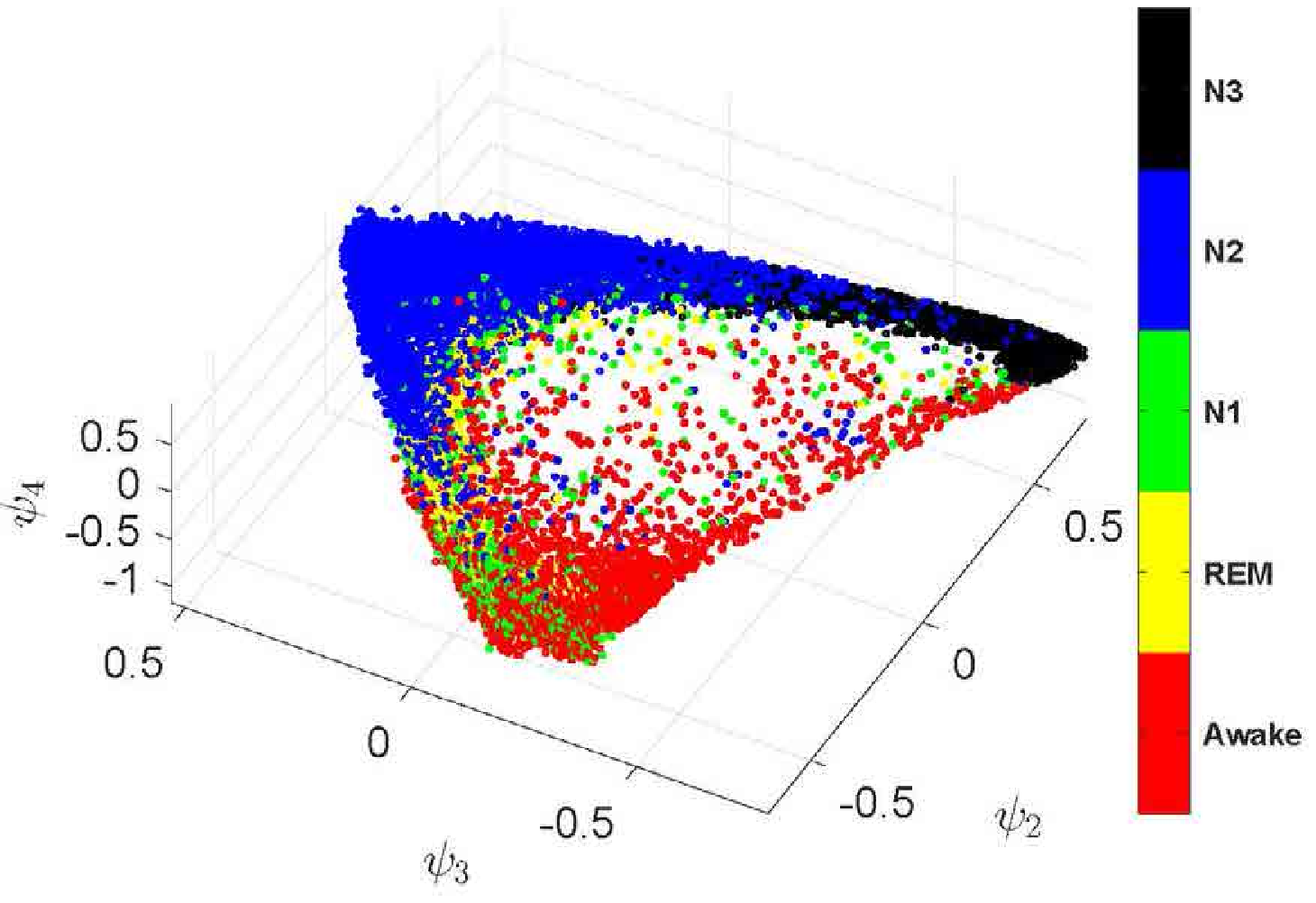}}\\
\subfigure{\includegraphics[scale=0.48]{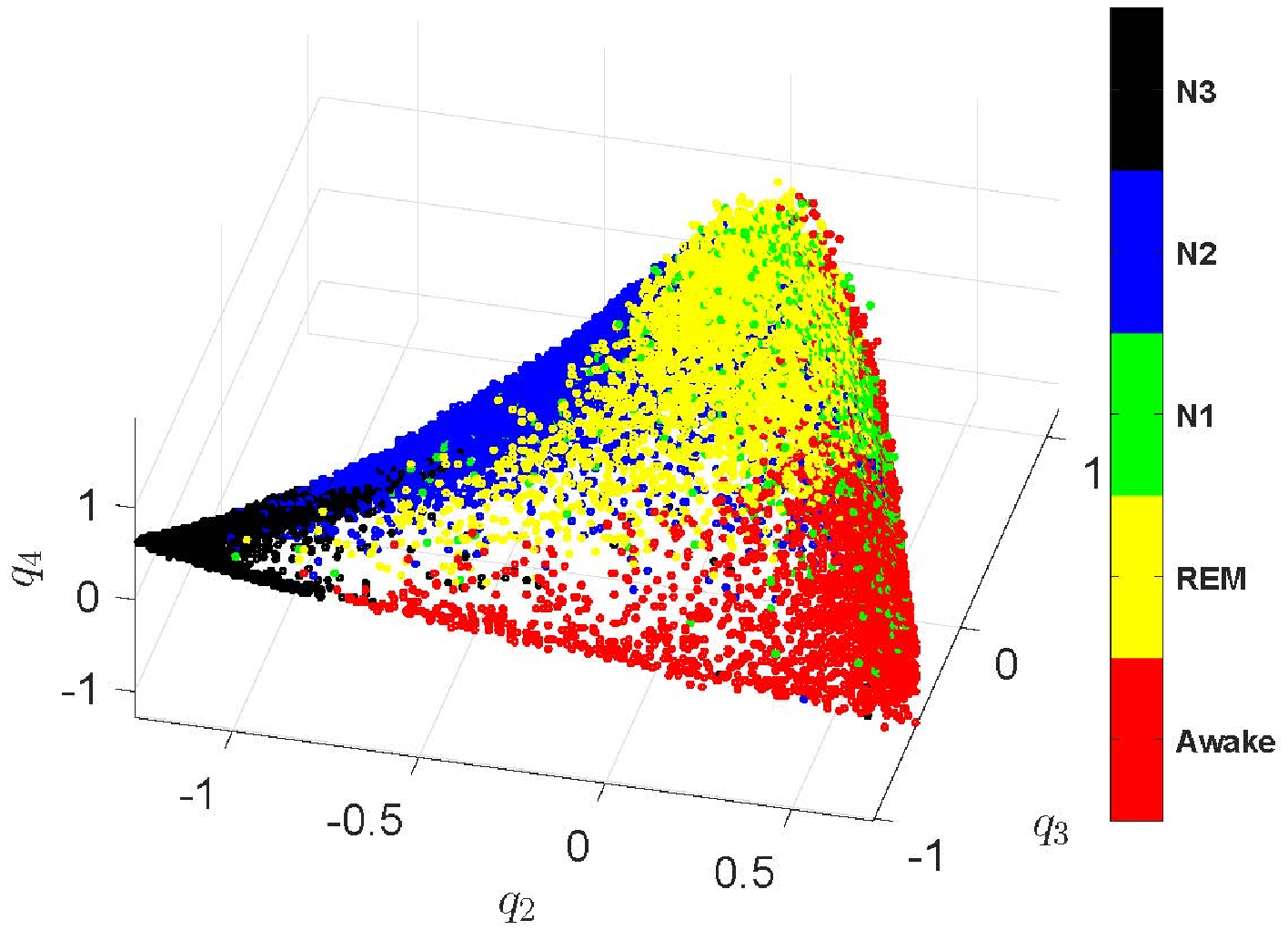}}\\
\subfigure{\includegraphics[scale=0.48]{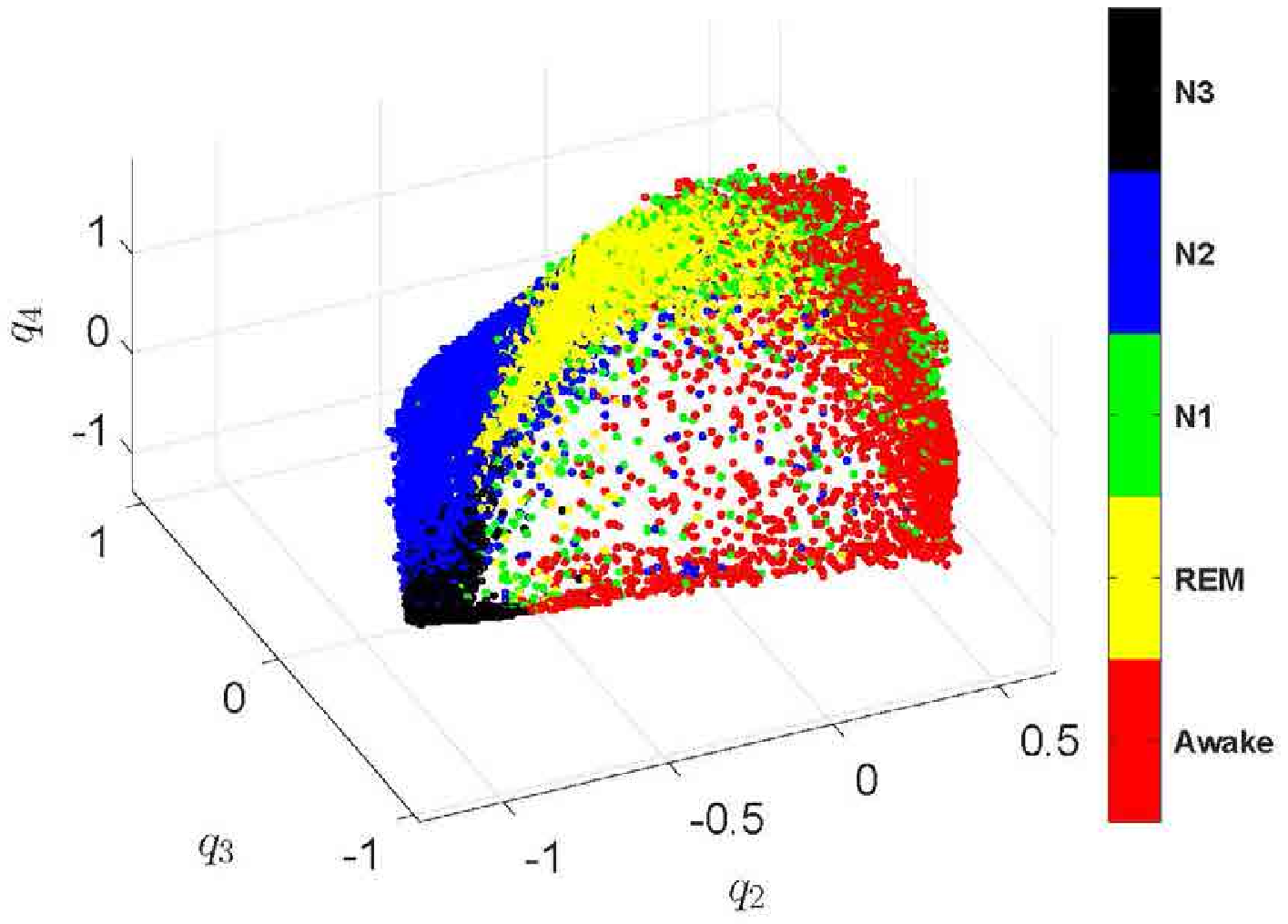}}
%\end{minipage}
%\begin{minipage}{0.3\textwidth}
\caption{A visualization of the common intrinsic sleep features (from two channels) extracted from 12 different subjects from the Sleep-EDF database (SC*). From top to bottom are ADM of Fpz-Cz \& Pz-Oz, multiview DM of Fpz-Cz \& Pz-Oz and multiview DM of Fpz-Cz \& Pz-Oz.
In the middle subplot, we plot $\{[q_2(i),q_3(i),q_4(i)]\}_{i=1}^J$, and in the bottom subplot, we show $\{[q_2(i+J),q_3(i+J),q_4(i+J)]\}_{i=1}^J$.
The ratios of the stages Awake, REM, N1, N2, and N3 are
17.3\%, 18.5\%, 4.3\%, 46.0\%, and 13.9\% respectively. Each point corresponds to a 30-second epoch.\label{fig:SCvisulization}
}
%\end{minipage}
\end{figure}

\begin{figure}[!htb]
\centering
\includegraphics[scale=0.26]{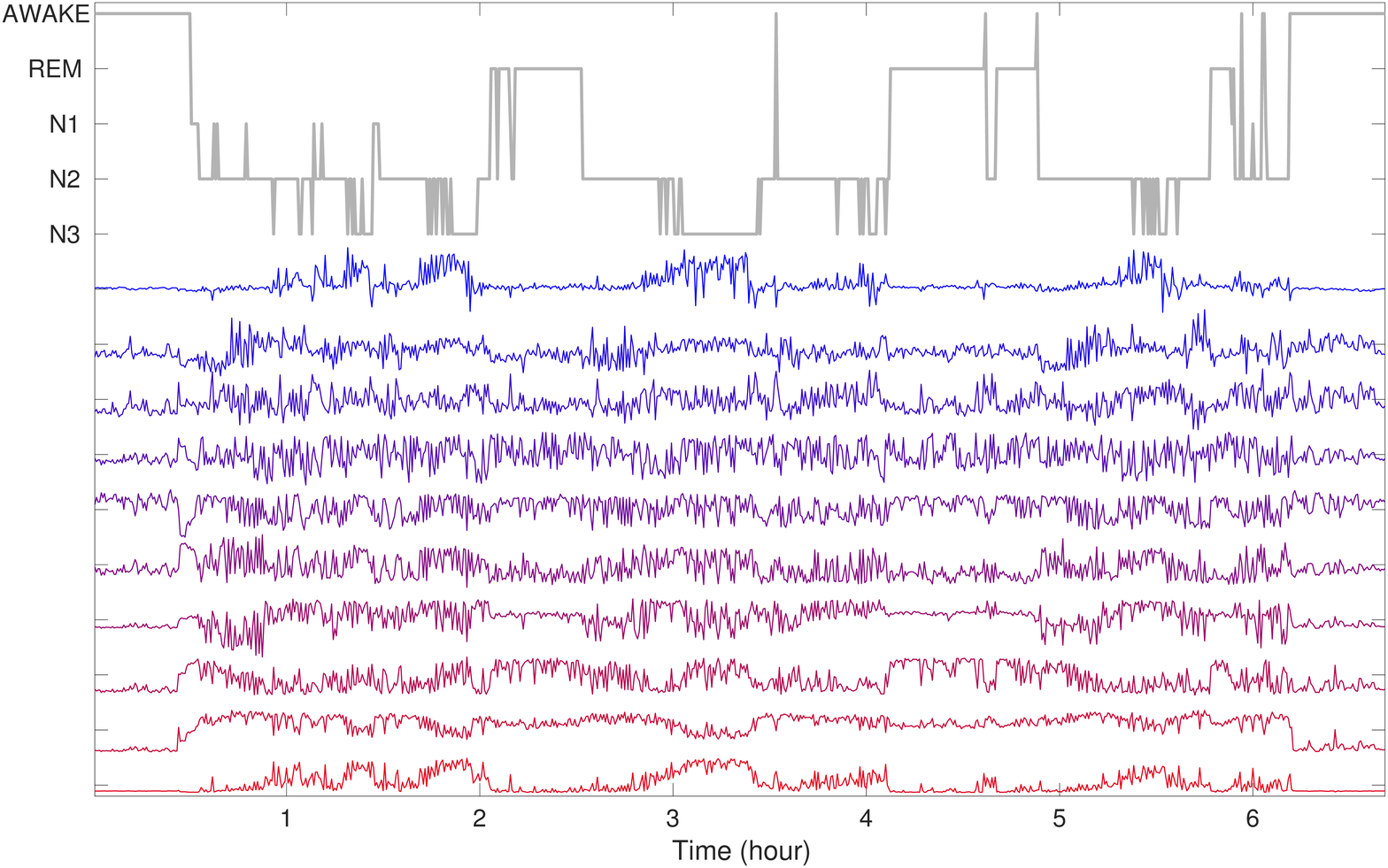}
\caption{A visualization of the final ten intrinsic features by ADM. The first is colored in red, and the tenth is colored in blue. The expert's labels are plotted on the top.}
\label{fig:Dynamics}
\end{figure}

We start from showing the visualization of the intrinsic sleep features and the common intrinsic sleep features from 12 different subjects in the SC* database. See Figure \ref{fig:SCvisulization0qq} for a visualization by DM. See Figure \ref{fig:SCvisulization} for a visualization by AD and co-clustering. Clearly, we see that Awake, REM, N2 and N3 stages are well clustered in all plots, while N1 is less clustered and tends to mixed up with other stages. Moreover, in AD and co-clustering, we further see a ``circle'' with a hole in the middle, and the sleep stages are organized on the circle and follow the usual sleep dynamics pattern.
While the geometric organization of sleep dynamics can be easily visualized in Figure \ref{fig:SCvisulization}, it is not easy to visualize the temporal dynamics information. For this purpose, we show the final intrinsic features expanded in the time line in Figure \ref{fig:Dynamics}. Another way to visualize the dynamics is via a video that encodes the temporal relationship among different sleep stages. See the video available in \url{https://www.dropbox.com/s/21e8aw7scvo5kkb/dynamics_SC31.mp4?dl=0}.

\begin{figure}
\centering
%\begin{minipage}{0.68\textwidth}
\subfigure
{\hspace{-0.95cm}\includegraphics[scale=0.48]{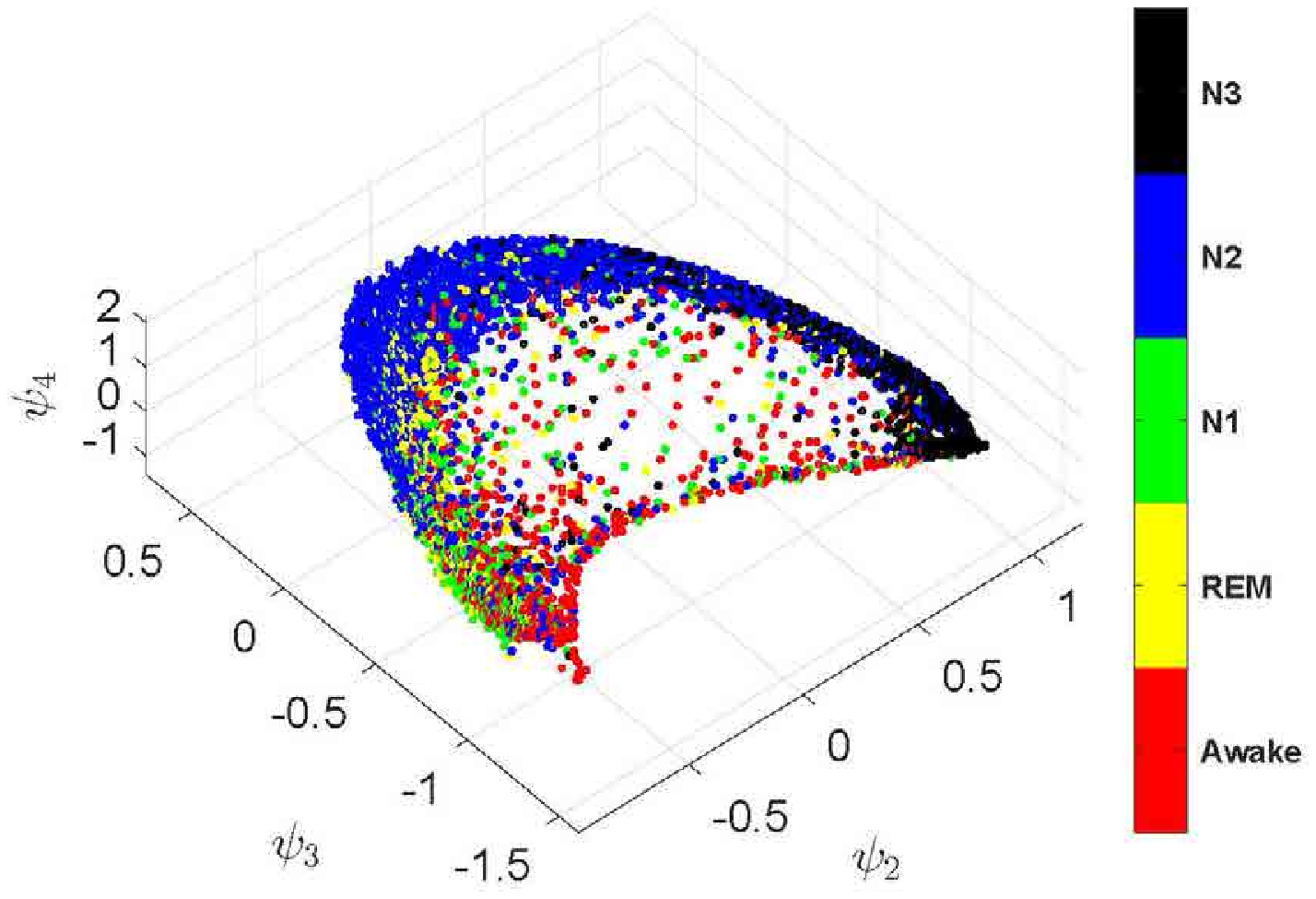}\label{Figure:2:ADM_ST}}
\subfigure
{\hspace{-0.95cm}\includegraphics[scale=0.48]{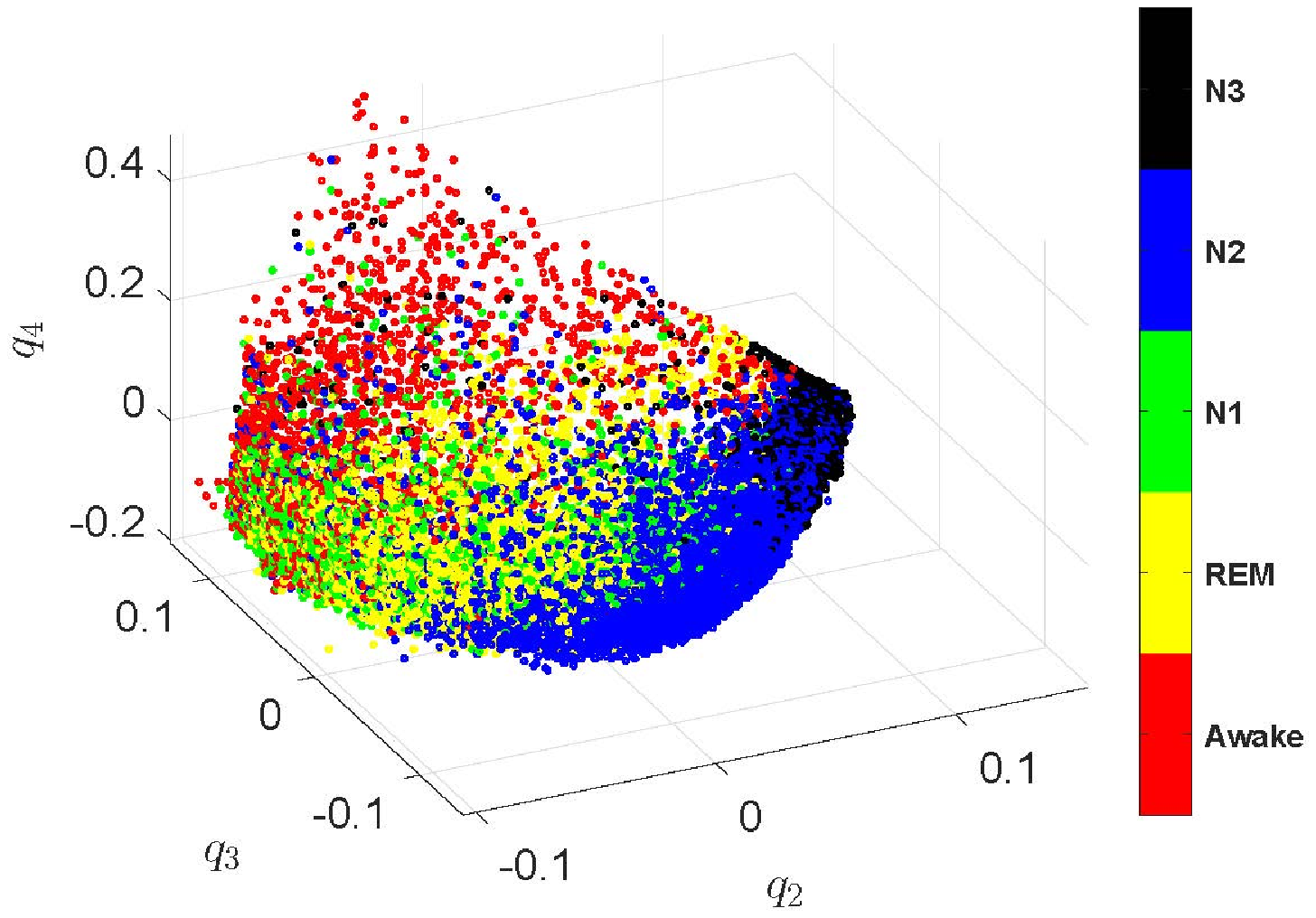}\label{Figure:2:mvDM1_ST}}
\subfigure
{\hspace{-0.95cm}\includegraphics[scale=0.48]{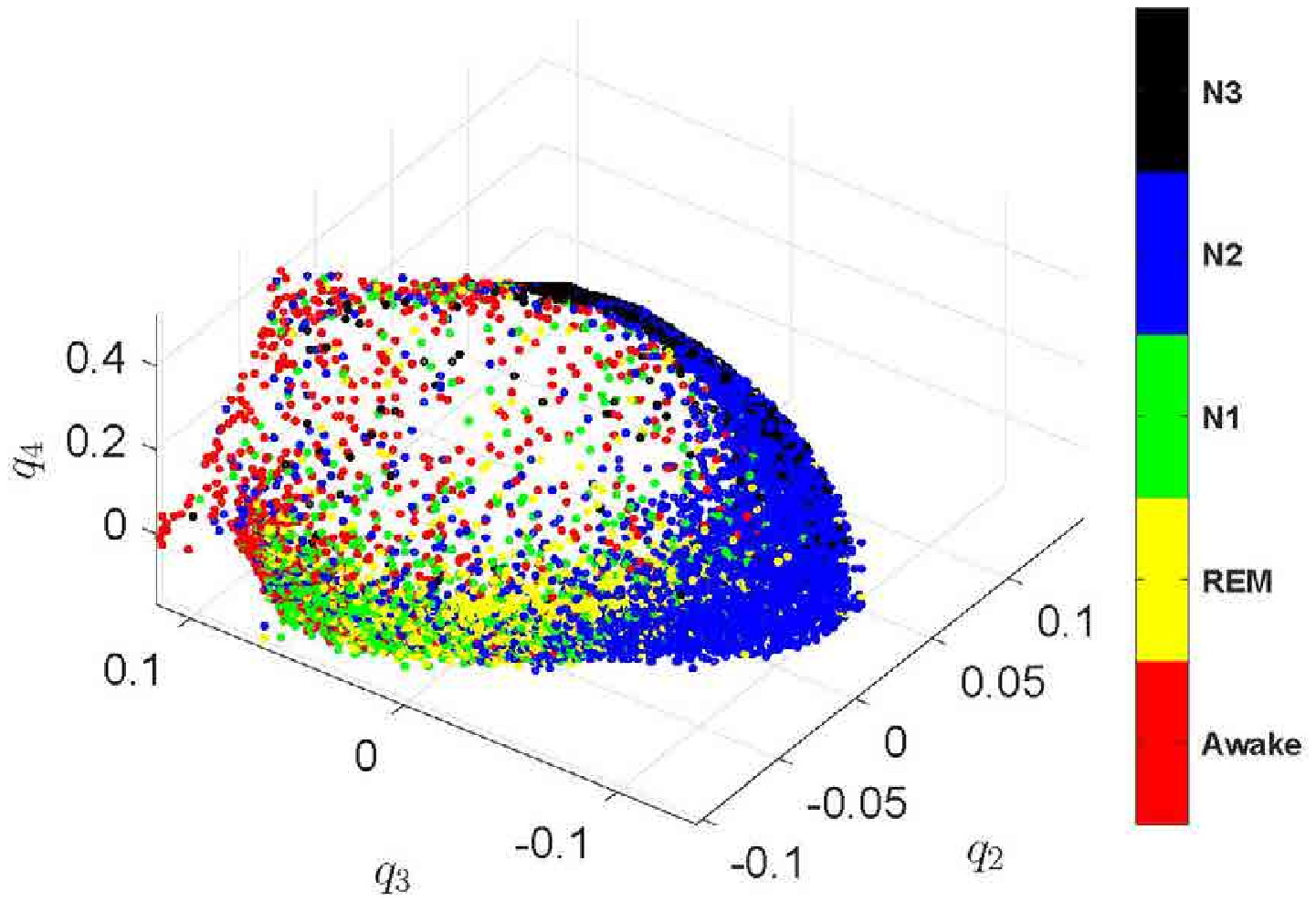}\label{Figure:2:mvDM2_ST}}
%\end{minipage}
%\begin{minipage}{0.3\textwidth}
\caption{A visualization of the common intrinsic sleep features (from two channels) extracted from 12 different subjects from the Sleep-EDF database (ST*). From top to bottom are ADM of Fpz-Cz \& Pz-Oz, first set of multiview DM of Fpz-Cz \& Pz-Oz and the second set of multiview DM of Fpz-Cz \& Pz-Oz.
In subplot \ref{Figure:2:mvDM1_ST}, we plot $\{[q_2(i),q_3(i),q_4(i)]\}_{i=1}^J$, and in subplot \ref{Figure:2:mvDM2_ST}, we show $\{[q_2(i+J),q_3(i+J),q_4(i+J)]\}_{i=1}^J$.
The ratios of the stages Awake, REM, N1, N2, and N3 are
10.3\%, 20.0\%, 10.0\%, 45.3\%, and 14.3\% respectively.  Each point corresponds to a 30-second epoch.
}
\label{fig:STvisulization}
%\end{minipage}
\end{figure}

Next, see Figure \ref{fig:STvisulization} for a visualization of AD and co-clustering of the ST* database. While Awake, REM, N2 and N3 stages are still well clustered in all plots, compared with the normal subjects in SC* database, the separation and the ``circle'' are less clear.

\subsection{Sleep stage prediction}
\begin{figure}[!htb]
\centering
\includegraphics[scale=0.3]{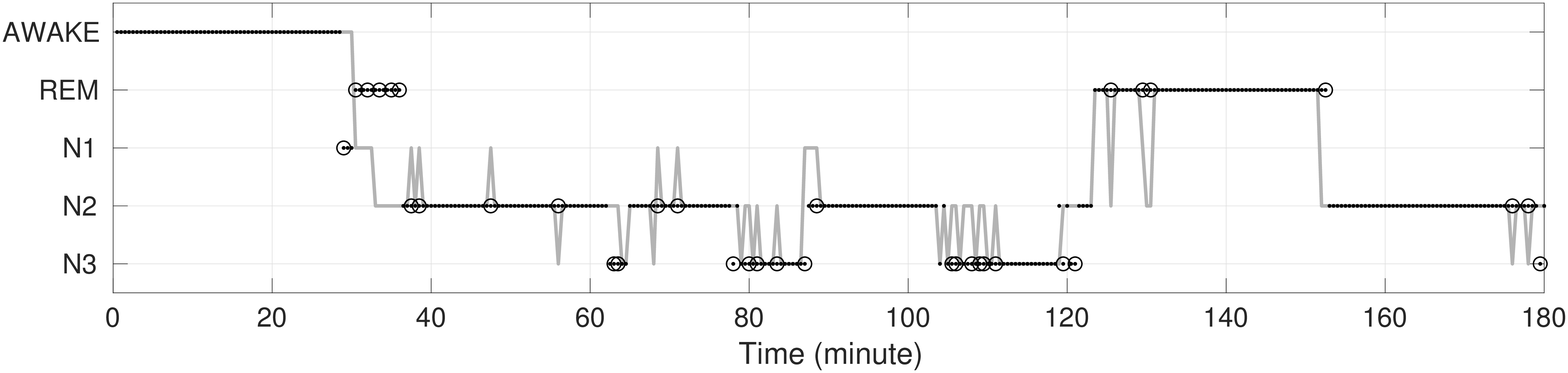}
\includegraphics[scale=0.3]{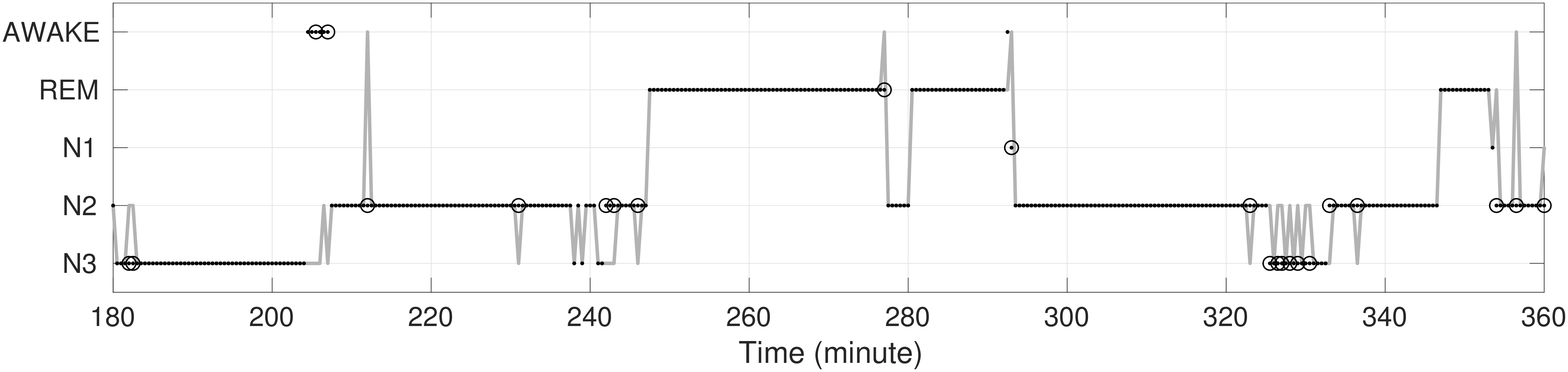}
\caption{The resulting hypnogram of one subject from SC*. The gray curve is the expert's label, and the black dots are the predicted sleep stages. The discrepancy is emphasized by the black circles.}
\label{fig:Hypnogram}
\end{figure}

Since there are long periods of wakefulness at the start and the end of recordings, when a subject is not sleeping, \cite{DeepSleepNet} only includes 30 minutes of such periods just before and after the sleep periods.
To have a fair comparison, we also follow this truncation rule.
In the end, the labeled epochs are imbalanced, with 42.4\% epochs labeled N2 and only 6.6\% epochs labeled N1. Authors of \cite{DeepSleepNet}, as well as \cite{2016autoencoder,tsinalis2016automatic}, handle the imbalanced data issue by setting the number of epochs per-stage per recording belonging the training data (consisting of ${\hat{K}}$ subjects) equal to the number of epochs of the least represented stage, N1. To have a fair comparison and test the stability of the proposed algorithm, we apply the same scheme in the following way.
We run the LOSOCV with ${\hat{K}}=9$. {For each subject in the testing set, we take {the} nine subjects with closest age and handle the imbalanced data like that of \cite{DeepSleepNet}.

The averaged confusion matrix of the proposed algorithm over 20 subjects is shown in Table \ref{table:SC2}.
The overall accuracy over 20 subjects is $82.49\%\pm 5.05\%$ and the macro F1 is $75.7\%\pm 5.2\%$, with Cohen's kappa $0.758\pm 0.07$. Note that the N1 prediction accuracy is the lowest one, with $35\%$ accuracy} compared with other stages, and most N1 epochs are classified as REM or N2.
This misclassification is related to the scattered N1 epochs in Figure \ref{fig:SCvisulization} that can be visually observed, and it is the main reason to drag down the overall accuracy and macro F1. We also note that N3 is commonly classified wrongly as N2, Awake is commonly classified wrongly as N1, and REM is commonly classified wrongly as N2.
{To further examine the performance, the resulting hypnogram of one subject is shown in  Figure \ref{fig:Hypnogram}. Note that the discrepancy between the experts' annotations and the prediction frequently happens when there is a ``stage transition''. Note that the sleep dynamics transition from one stage to another one often happens in the middle of one epoch. Thus, those epochs with sleep dynamics transition contain information that is not purely for one stage, and hence harder to classify.}

\begin{table}
\scriptsize
\setlength\extrarowheight{3pt}
\caption{Comparison matrix obtained from 20-fold leave-one-subject-out cross-validation on Fpz-Cz and Pz-Oz channels from the Sleep-EDF SC* database. The common intrinsic sleep feature is used. The overall accuracy equals 82.57\%, the macro F1 score equals 76.0\% and Cohen's kappa equal 0.763. If the classification accuracy, macro F1 score, and Cohen's kappa are computed for each night recording, the standard deviation of classification accuracy (resp. the macro F1 score and Cohen's kappa) for the 39-night recordings is 4.96\% (resp. 5.15\% and 0.068).
We follow the class-balanced random sampling scheme used in \cite{2016autoencoder,tsinalis2016automatic,DeepSleepNet}.}
\centering
\begin{tabular}{|c|ccccc|ccc|}
\toprule
 &\multicolumn{5}{c}{\bf Predicted} & \multicolumn{3}{|c|}{\bf Per-class Metrics}\\
  & \multicolumn{1}{c}{\centering Awake} & \multicolumn{1}{c}{\centering REM} &  \multicolumn{1}{c}{\centering N1} & \multicolumn{1}{c}{\centering N2} & \multicolumn{1}{c}{\centering N3} & \multicolumn{1}{|c}{\centering PR} & \multicolumn{1}{c}{\centering RE} & \multicolumn{1}{c|}{\centering F1} \\
\midrule
Awake ({18}\%) &  6943 ({\bf 88\%})  &     184  ({\bf 2\%})    &    625 ({\bf 8\%})  &    156  ({\bf2\%})     &    19 ({\bf 0\%}) &  91    & 88  & 89 \\
REM   ({18}\%) &  112 ({\bf 1\%})    &     7063  ({\bf92 \%}) &      123  ({\bf 2\%})  &    419  ({\bf 5\%})    &    0 ({\bf 0\%}) &  73  &  92 &  81 \\
N1    ({7}\%)  &   378 ({\bf 13\%})   &      907  ({\bf 32\%})  &    967  ({\bf 35\%})  &      534  ({\bf 19\%})   &   18 ({\bf1 \%}) &   45  & 34   &  39 \\
N2    ({42}\%) &  128 ({\bf 1\%})    &     1451  ({\bf 8\%})    &    412    ({\bf 2\%})   &       14557 ({\bf82\%})  &   1251 ({\bf7 \%}) &   90 &  82  &  86  \\
N3    ({14}\%) &  29 ({\bf 0\%})    &      16 ({\bf 0\%})      &   3  ({\bf 0\%})       &     545   ({\bf10 \%})   &   5110 ({\bf 90\%})  &  80 &  90   &  84  \\
\bottomrule
\end{tabular}
\label{table:SC2}
\end{table}

{An ideal approach to handle the imbalanced data is collecting more data to enhance the prediction accuracy. In the SC${}^*$ dataset,
there were long periods of awake epochs before the start and after the end of sleep that we can use. To further evaluate the algorithm, we consider longer periods of wakefulness just before and after the sleep periods.
Apart from the 3 recordings (sc4092e0, sc4191e0, sc4192e0), we included 90 minutes of awake periods before and after the sleep periods.
For the sc4092e0, sc4191e0, and  sc4192e0 recordings, we only included 60 minutes of awake periods just before and after the sleep periods due to the appearance of artifacts (labeled as MOVEMENT and UNKNOWN), which were at the start or the end of each recording. With more awake epochs, the corresponding comparison matrix with the same  is shown in
Table \ref{table:SC_moreAwake}.
All performance indices, including the overall accuracy, the macro F1 score and the Cohen's kappa,
are consistently higher than those from only including 30 minutes of awake periods
just before and after the sleep periods reported in Table \ref{table:SC2}.
The overall accuracy over 20 subjects is $84.21\%\pm 4.85\%$ and the macro F1 is $76.5\%\pm 5.18\%$, with Cohen's kappa $0.79\pm 0.06$.
Particularly, the accuracy of N1 prediction is increased to 42\%.

\begin{table}[h]
\scriptsize
\setlength\extrarowheight{3pt}
\caption{{Confusion matrix obtained from 20-fold leave-one-subject-out cross-validation on Fpz-Cz and Pz-Oz channels from the Sleep-EDF SC* database with longer awake periods. The common intrinsic sleep feature is used. The overall accuracy equals 84.21\%, the macro F1 score equals 76.5\% and Cohen's kappa equal 0.788. If the classification accuracy, macro F1 score, and Cohen's kappa are computed for each night recording, the standard deviation of classification accuracy (resp. the macro F1 score and Cohen's kappa) for the 39-night recordings is 4.85\% (resp. 5.18\% and 0.06).
We apply the class-balanced random sampling scheme proposed in \cite{2016autoencoder,tsinalis2016automatic,DeepSleepNet}.}}
\centering
\begin{tabular}{|c|ccccc|ccc|}
\toprule
 &\multicolumn{5}{c}{\bf Predicted} & \multicolumn{3}{|c|}{\bf Per-class Metrics}\\
  & \multicolumn{1}{c}{\centering Awake} & \multicolumn{1}{c}{\centering REM} &  \multicolumn{1}{c}{\centering N1} & \multicolumn{1}{c}{\centering N2} & \multicolumn{1}{c}{\centering N3} & \multicolumn{1}{|c}{\centering PR} & \multicolumn{1}{c}{\centering RE} & \multicolumn{1}{c|}{\centering F1} \\
\midrule
Awake ({34}\%) &  15159 ({\bf 88\%})  &    339   ({\bf 2\%})    &    1572 ({\bf 9\%})  &    170  ({\bf1\%})     &    45 ({\bf 0\%}) &  98    & 88  & 92 \\
REM   ({15}\%) &  24 ({\bf 1\%})    &     7162  ({\bf93 \%}) &      133  ({\bf 2\%})  &    395  ({\bf 5\%})    &    3 ({\bf 0\%}) &  75  &  93 &  83 \\
N1    ({5}\%)  &   232 ({\bf 8\%})   &      829  ({\bf 30\%})  &    1180   ({\bf 42\%})  &      544  ({\bf 19\%})   &   19 ({\bf1 \%}) &   33  & 42   &  37 \\
N2    ({35}\%) &  89 ({\bf 0\%})    &     1196  ({\bf 7\%})    &    636    ({\bf 4\%})   &       14553 ({\bf82\%})  &   1325 ({\bf7 \%}) &   90 &  82  &  86  \\
N3    ({11}\%) &  19 ({\bf 0\%})    &      3 ({\bf 0\%})      &   21  ({\bf 0\%})       &     499   ({\bf9 \%})   &   5161 ({\bf 91\%})  &  79 &  91   &  84  \\
\bottomrule
\end{tabular}
\label{table:SC_moreAwake}
\end{table}

For the ST* database, we also run the LOSOCV with $\hat{K}=9$. The averaged confusion matrix of the proposed algorithm over 22 subjects is shown in Table \ref{table:ST}.
The overall accuracy is $77.8\%\pm 5.77\%$ and the macro F1 is $71.5\%\pm 7.55\%$, with Cohen's kappa $0.69\pm 0.08$. In this database, there are $10\%$ epochs labeled as N1. Although it is slightly higher than that of SC$^{*}$ database, the prediction performance of N1 is $34\%$, which is still relatively low. Also, note that the prediction performance of N3 is lower, and a significant portion of N3 is mis-classified as N2.

\begin{table}
\scriptsize
\setlength\extrarowheight{6pt}
\caption{Comparison matrix obtained from 22-fold leave-one-subject-out cross-validation on Fpz-Cz and Pz-Oz channels from the Sleep-EDF ST$^{*}$ database. The common intrinsic sleep feature is used. The overall accuracy equals 77.01\%,
the macro F1 score equals 71.53\%, and
Cohen's kappa equals 0.6813.
The standard deviation of classification accuracy (resp., the macro F1 score and Cohen's kappa) for the 22-night recordings is 6.63\%
(resp., 7.78\% and 9.27\%). We apply the class-balanced random sampling scheme proposed in \cite{2016autoencoder,tsinalis2016automatic,DeepSleepNet}.}
\centering
\begin{tabular}{|c|ccccc|ccc|}
\toprule
 &\multicolumn{5}{c}{\bf Predicted} & \multicolumn{3}{|c|}{\bf Per-class Metrics}\\
  & \multicolumn{1}{c}{\centering Awake} & \multicolumn{1}{c}{\centering REM} &  \multicolumn{1}{c}{\centering N1} & \multicolumn{1}{c}{\centering N2} & \multicolumn{1}{c}{\centering N3} & \multicolumn{1}{|c}{\centering PR} & \multicolumn{1}{c}{\centering RE} & \multicolumn{1}{c|}{\centering F1} \\
\midrule
Awake (11\%) &  2008 ({\bf 88\%}) &   40 ({\bf 2\%})    &  178  ({\bf 8\%})  &  42 ({\bf 2\%})     & 16  ({\bf 0\%})   &   72   & 88  &   79 \\
REM (20\%)   & 59 ({\bf 1\%})     &   3489 ({\bf 85\%}) &   238  ({\bf 6\%}) &  334  ({\bf 8\%})  & 11  ({\bf 0\%})    &  79   &  84 &  81 \\
N1  (10\%)   & 548  ({\bf 27\%})  &   388  ({\bf 19\%}) &   697  ({\bf 34\%}) &  409 ({\bf 20\%})   &  2 ({\bf 0\%})     &   48  &  34  &  40 \\
N2   (45\%)  &  155 ({\bf 2\%})   &   514 ({\bf 5\%})  &  348  ({\bf 4\%})  &  7599  ({\bf 80\%}) &  877 ({\bf 9\%})   &  84   & 80  &  82  \\
N3   (15\%)  & 30  ({\bf 1\%})    &     2  ({\bf 0\%})  &   5 ({\bf 0\%})    &  654  ({\bf 21\%})  &  2454  ({\bf 79\%})  &  73   &   78 &  75  \\
\bottomrule
\end{tabular}
\label{table:ST}
\end{table}

\subsection{More comparisons for sleep stage prediction}
{To appreciate the significance of the diffusion geometry based sensor fusion framework, we report the results without two critical setups in the proposed algorithm -- the sensor fusion and the local MD.
First, we consider the case if we simply concatenate intrinsic sleep features of two channels, instead of taking the common intrinsic sleep features; that is, we concatenate $\Phi^x_t(\mathbf{u}^{(j)})$ and $\Phi^y_t(\mathbf{u}^{(j)})$ in \eqref{Definition:intrinsic sleep features} directly to replace \eqref{Definition:common intrinsic sleep feature} when we train the HMM model.
Second, we consider the case if we do not use local MD to compare synchrosqueezed EEG spectral features but the ordinary $L^2$ distance; that is, $d^{2}(\mathbf{u}^{(i)},\mathbf{u}^{(j)})$ in \eqref{affinity_matrix} is defined as $\|\mathbf{u}^{(i)}-\mathbf{u}^{(j)}\|_{\mathbb{R}^{10}}$ instead of $d_{\texttt{LMD}}^{2}(\mathbf{u}^{(i)},\mathbf{u}^{(j)})$.
Third, we consider the single EEG channel; that is, we run HMM on the intrinsic sleep features extracted from Fpz-Cz or Pz-Oz channel.

The results of the above three combinations (confusion matrices not shown) for the SC${}^*$ database are shown in Figure \ref{fig:effect_AD}. Note that the mean and standard deviation of ACC, MF1 and Cohen's kappa are evaluated from all subjects, which are different from that shown in Table \ref{table:SC2}.
It is clear that the averaged ACC, MF1 and Cohen's kappa are consistently downgraded in these three cases.
In Figure \ref{fig:effect_AD}, we see that compared with single channel or sensor fusion without local MD, the proposed sensor fusion of two channels consistently improve the result with statistical significance, for both mean and variance of ACC and MF1. Although we can see a significant difference between the mean and variance of Cohen's kappas before the Bonfferoni correction in these comparisons, the significance is not strong enough to stand the Bonfferoni correction.
Compared with a direct concatenation of intrinsic sleep features of two channels, we see that there is a significance between the averaged ACCs before the Bonferroni correction, but it disappears after the Bonferroni correction. However, we see a significant difference between the variance of ACC and MF1.
This fact reflects the essential property of the diffusion-based unsupervised learning algorithms. Via the alternating diffusion process for the sensor fusion, the intrinsic sleep features are ``stabilized'', and hence the smaller variance.
This comparison provides an empirical evidence of the usefulness of the sensor fusion and local Mahalanobis distance in the proposed algorithm, in addition to the established theoretical backup shown in the Online Supplementary.

\begin{figure}
\centering
\subfigure[][Comparison of accuracy]
{\includegraphics[scale=0.3]{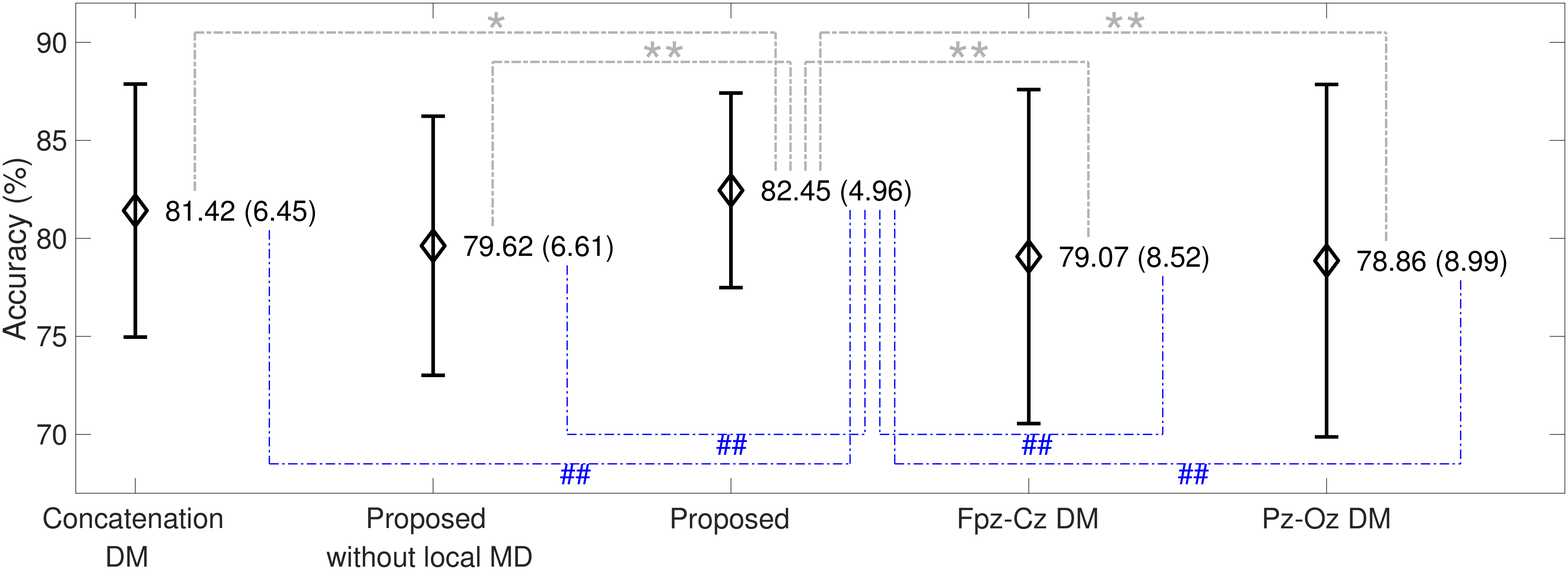}}\\
\subfigure[][Comparison of macro-F1]
{\includegraphics[scale=0.3]{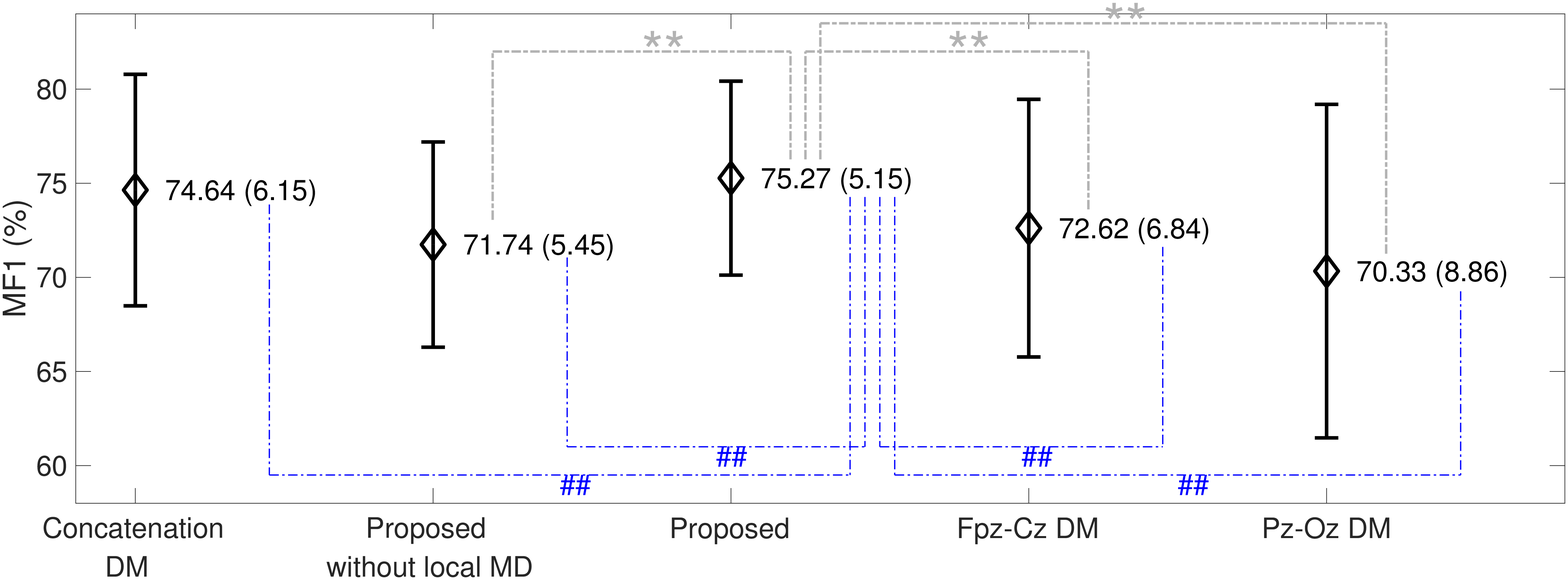}}\\
\subfigure[][Comparison of Cohen's kappa]
{\includegraphics[scale=0.3]{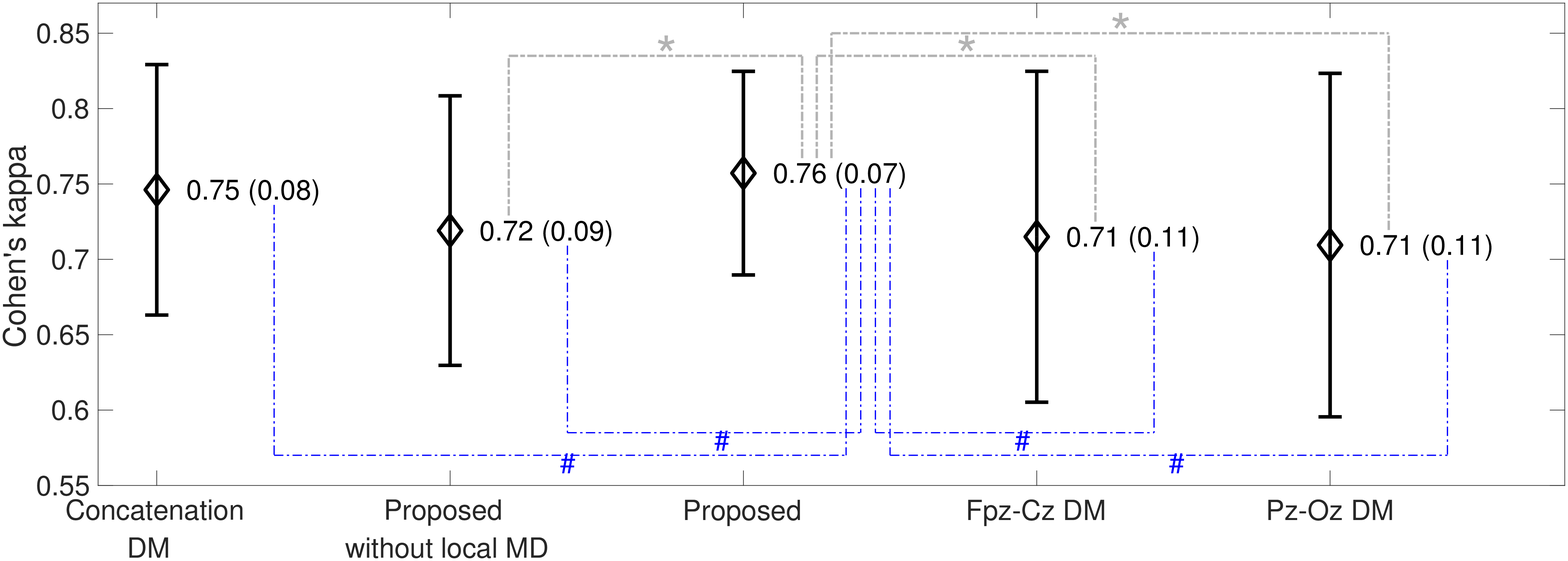}}
\caption{{Comparison between different information fusion methods in terms of the accuracy (ACC), macro F1-score (MF1) and Cohen's Kappa for the SC* database. To evaluate if the mean is improved, the one-tail Wilcoxon signed-rank test is applied under the null hypothesis that the difference between the pairs follows a symmetric distribution around zero.
To evaluate if the variance is smaller, we apply the one-tail F test under the null hypothesis that there is no difference between the variances.
$\star$ (respectively $\star\star$) means statistical significance without (respectively with) the Bonferroni connection when the mean is compared; $\#$ (respectively $\#\#$) means statistical significance without (respectively with) the Bonferroni connection when the variance is compared.}
}
\label{fig:effect_AD}
\end{figure}

We have considered LOSOCV to prevent overfitting. Here we further consider the 5-fold leave-subject-out cross validation (CV) to further evaluate the proposed algorithm; that is, we randomly divide all subjects into 5 non-overlapping groups, and for each group as the testing group, we train the model from the left four groups. The model is trained in the following way. For each subject in the testing set, we take epochs of {the} nine ($\hat{K}=9$) subjects with closest age from the other four groups collected during the first night into account, and balance classes by taking the epochs from the second night.
The result is reported in Table \ref{table:SC_5cv}. We see that the result with 5-fold leave-subject-out CV is similar to the leave-one-subject-out CV.

\begin{table}
\scriptsize
\setlength\extrarowheight{3pt}
\caption{Comparison matrix obtained from 5-fold cross-validation on Fpz-Cz and Pz-Oz channels from the Sleep-EDF SC* database. The common intrinsic sleep feature is used. The 20 subjects in the SC* database are divided into 5 groups. Each group contains 4 subjects. After one of the 5 groups is selected for testing, the remaining 4 groups are used for training purposes. The overall accuracy equals 82.25\%, the macro F1 score equals 75.89\% and Cohen's kappa equal 0.7591. If the classification accuracy, macro F1 score, and Cohen's kappa are computed for each night recording, the standard deviation of classification accuracy (resp. the macro F1 score and Cohen's kappa) for the 39-night recordings is 4.99\% (resp. 5.05\% and 6.68\%).
The training set consist of two-night recordings of the remaining 19 subjects with class-balanced random sampling, which is the same as the class balancing method used in \cite{2016autoencoder,tsinalis2016automatic,DeepSleepNet}.}
\centering
\begin{tabular}{|c|ccccc|ccc|}
\toprule
 &\multicolumn{5}{c}{\bf Predicted} & \multicolumn{3}{|c|}{\bf Per-class Metrics}\\
  & \multicolumn{1}{c}{\centering Awake} & \multicolumn{1}{c}{\centering REM} &  \multicolumn{1}{c}{\centering N1} & \multicolumn{1}{c}{\centering N2} & \multicolumn{1}{c}{\centering N3} & \multicolumn{1}{|c}{\centering PR} & \multicolumn{1}{c}{\centering RE} & \multicolumn{1}{c|}{\centering F1} \\
\midrule
Awake (18\%) &  6857 ({\bf 86\%})  &    180   ({\bf 2\%})    &    732 ({\bf 9\%})  &    148  ({\bf2\%})     &    10 ({\bf 0\%}) &  92    & 87  & 89 \\
REM   (18\%) &  124 ({\bf 2\%})    &     6965  ({\bf90 \%}) &      184  ({\bf 2\%})  &    443  ({\bf 6\%})    &    1 ({\bf 0\%}) &  75  &  90 &  82 \\
N1    (7\%)  &   327 ({\bf 12\%})   &      873  ({\bf 31\%})  &    1032   ({\bf 37\%})  &      550  ({\bf 20\%})   &   22 ({\bf0 \%}) &   42  & 37   &  39 \\
N2    (42\%) &  131 ({\bf 1\%})    &     1301  ({\bf 7\%})    &    528    ({\bf 3\%})   &       14517 ({\bf82\%})  &   1322 ({\bf7 \%}) &   90 &  82  &  85  \\
N3    (14\%) &  33 ({\bf 1\%})    &      13 ({\bf 0\%})      &   11  ({\bf 0\%})       &     512   ({\bf9 \%})   &   5134 ({\bf 90\%})  &  79 &  90   &  84  \\
\bottomrule
\end{tabular}
\label{table:SC_5cv}
\end{table}

}

\section{Discussion and conclusion}\label{Section:Conclusion}

{The unsupervised diffusion geometry based sensor fusion framework is proposed to capture the geometric structure of the sleep dynamics. We take the spectral information of EEG signals as an example and test the framework on the publicly available benchmark database.}
With the learning algorithm HMM, we obtain an accurate prediction model, and the result is compatible with several state-of-the-art algorithms {based on neural network (NN). In addition to the theoretical backup of the diffusion geometry framework provided in the online supplementary materials, a systematical examination of step step in the diffusion geometry framework is provided. All these summarize the usefulness of the diffusion geometry framework for the sleep dynamics study. We mention that the proposed framework is flexible to study other physiological dynamics but not only for studying the sleep dynamics. For example, its variation has been applied to study f-wave from subjects with atrial fibrillation \cite{Malik_Reed_Wang_Wu:2017}, intra-cranial EEG signal \cite{Alagapan_Shin_Frohlich_Wu:2018}, etc.}

\subsection{Physiological explanation of the results}
Although our overall prediction accuracy is compatible with the state-of-the-art prediction algorithm in the Sleep-EDF SC* database, like \cite{DeepSleepNet}, we see that the prediction accuracy of N1 is relatively low by our algorithm (F1 is 39\% by our method and 46.6\% in \cite{DeepSleepNet}), and this low N1 accuracy downgrades the overall accuracy. This low prediction rate of N1 is also seen in the Sleep-EDF ST* database. This low prediction rate partially comes from the relatively small size of N1 epochs, and partially comes from the algorithm and available channels.

Based on the AASM criteria \cite{Iber2007}, to distinguish N1 and REM, we need electrooculogram and electromyogram {signals}, which are not available in the dataset. The EEG backgrounds of N1 and N2 are the same, and experts distinguish N1 and N2 by the K-complex or spindle, as well as the {\em 3-minute rule}. While the synchrosqueezed EEG spectral features capture the K-complex or spindle behavior, the 3-minute rule is not considered in the algorithm.
In the proposed algorithm, in order to handle the inter-individual variability, the temporal information among epochs is not fully utilized {when we design the intrinsic sleep feature but only used in the HMM}.
How to incorporate the temporal information into the diffusion geometry framwork will be explored in the future.
{Furthermore, there are other information in addition to the spectral information discussed in this paper. We do not extensively explore all possible information, but focus on the diffusion geometry and sensor fusion framework.} For example, while the vertex sharp is a common ``landmark'' indicating transition from N1 to N2, {we do not take it into account since this feature is not always present and a rule-based approach is needed to include this temporally transient feature.
Another interesting reasoning that it is possible to improve the N1 accuracy is {the} deep neural network (DNN) result. This suggests that by taking experts' labels into account,} some distinguishable EEG structure of N1 that is not sensible by spectral information can be efficiently extracted by the DNN framework proposed in \cite{DeepSleepNet}. In conclusion, since the proposed features depend solely on the spectral information, we may need features of different categories to capture this unseen N1 structure.
On the other hand, it is well known that different EEG leads provide different information for N1. For example, the occipital lead has a stronger alpha wave strength, compared with the central lead, when transiting from wakefulness to N1. When there are multiple EEG leads, this lead information could be taken into account to further improve the accuracy.

Note that beside N1, the prediction performance of N3 is also lower in the Sleep-EDF ST* database, where the subjects take temazepam before data recording. It has been well known that in general benzodiazepine hypnotics \cite{Borbely1985} reduces the low frequency activity and enhances spindle frequency. Since our features are mainly based on the spectral information, a N3 epoch might look more like N2 epochs, and hence the confusion and the lower performance. This outcome emphasizes the importance of the drug history when {designing} the algorithm.

\subsection{Visualization for sleep dynamics exploration}
In Figures \ref{fig:SCvisulization} and \ref{fig:STvisulization}, we show the underlying geometric structure of the sleep dynamics captured by the proposed algorithm -- the {Awake}, REM, N2 and N3 are well clustered, with N1 scattered around {Awake}, REM, and N2.
Furthermore, in Figures \ref{fig:SCvisulization} and \ref{fig:STvisulization}, we can even visualize a ``circle''.
An interesting physiological finding from these plots is the close relationship between N3 and Awake. Note that the same result is also shown in Figure \ref{FlowChart}. This geometric relationship indicates the similarity between the common intrinsic sleep features of N3 and Awake stages. This similarity comes from the well known fact that before arousal, particularly across the descending part of sleep cycles and in the first cycle, we can observe the ``delta-like burst'' that mimics the delta wave specific for N3 stages \cite{BonnetCarley1992,Halasz2004}. Note that epochs from one subject are used to generate Figure \ref{FlowChart}, and 12 different subjects are pooled together to generate Figures \ref{fig:SCvisulization} {and} \ref{fig:STvisulization}, and we see the same geometric structure. This finding {exemplifies} our observation that this distribution is consistent across subjects. On the other hand, due to the sleep apnea disturbance, this ``circle'' is in general less obvious. This indicates the interruption of sleep dynamics by sleep apnea and hence the frequent random transition from one sleep stage to another. A possible direction is applying the topological data analysis tools to quantify the existence of circle, and hence a quantification of sleep apnea disturbance.

\subsection{Sleep stage classification and comparison with related work}
\label{sec:relatedwork}

There have been many proposed algorithms for the sake of automatic sleep stage scoring.
Since we focus on the LOSOCV scheme and predict five different sleep stages, here we only mention papers considering the LOSOCV scheme and predicting five different sleep stages from single- or two- EEG channels.

In \cite{2016autoencoder}, the performance of the stacked sparse autoencoder was evaluated in Sleep-EDF SC$^{*}$, and the overall accuracy was 78.9\%.
Instead of extracting features based on the domain knowledge, features in \cite{tsinalis2016automatic} are automatically learned by the convolutional neural networks (CNNs). The overall accuracies was 74.8\% for the Sleep-EDF SC$^{*}$ database.
In \cite{DeepSleepNet}, the authors proposed a deep learning model, called DeepSleepNet, which reaches the state-of-the-art 82.0\% of overall accuracy on the Sleep-EDF SC$^{*}$ database. In \cite{Vilamala2017}, a similar approach based on the deep CNN with modifications is considered and achieves a compatible result. All the above studies focus on the single channel EEG signal.

Compared with the state-of-the-art DNN approach \cite{DeepSleepNet}, which is supervised in nature, our approach is unsupervised in nature. Recall that the main difference between the supervised learning and unsupervised learning is that the label information is taken into account in the supervised learning approach. The success of DNN in many fields is well known \cite{LeCun2015}, and it is not surprising that it has a great potential to help medical data analysis.

{While DNN is in general an useful tool for the engineering purpose, it is often criticized {of working as a black box}. For medical problems and datasets, when interpretation is needed, a mathematically solid and interpretable tool would be useful.
The algorithm we proposed, on the other hand, has a clear interpretation with solid mathematical supports.}
Moreover, a peculiar property of medical databases, the ``uncertainty'', deserves more discussion.
Take the sleep dynamics studied in this paper as an example. It is well known that the inter-expert agreement rate is only about 80\% for normal subjects, not to say for subjects with sleep problems \cite{Norman2000}. With this uncertainty, a supervised learning algorithm {\em might} learn both good and bad labels. On the other hand, the proposed unsupervised approach is independent of the provided labels, and the chosen {spectral} features all come from the {EEG signal}, and speak solely for the {sleep dynamics} but not the labels. To some extent, the ``uncertainty'' issue is less critical via the unsupervised approach, since the uncertain labels are not taken into account in the feature extraction step.

Since both supervised and unsupervised approaches have their own merits, it is natural to seek for a way to combine both. We are exploring the possibility of combining DNN and the proposed unsupervised techniques, and the result will be reported in the future work.

\subsection{``Self-evolving'' artificial intelligence system}

Due to the advance of the technology and computational power, in the past decades a lot of effort has been devoted to establish an artificial intelligence (AI) system for the automatic sleep stage annotation purpose. In addition to being able to accurately score sleep stages, an ideal AI system should also be able to ``self-learn'' or accumulate knowledge from the historical database.
Note that despite the well accepted homeostasis assumption of physiological system, physiological dynamics vary from subject to subject. Therefore, the main challenge of this self-learning capability is handling the inter-individual variability. This challenge is actually ubiquitous -- for a new-arriving subject, {how may one utilize} the existing database with the expert annotations?

This challenge is actually empirically handled in this article. Recall that for each given subject, we take the age into account to find ``similar subjects'' to train the model to automatically annotate the sleep stage of the given subject.
This idea is a special case of the general picture commonly encountered in the clinical situation. The inter-individual variability is inevitable, and this variability is the main obstacle toward a self-evolving system that can self-learn like a human being. Our solution is respecting the physicians' decision making process and clinical experience to build up the system -- when a physician sees a subject the first time, an automatic ``classification'' of this subject is established. This ``decision tree'' is mostly based on the physician's knowledge. Although it varies from physician to physician, the overall structure of the decision tree ``should be'' relatively stable, ideally. For example, a physician will not consider any menopause-related diseases if the patient is a five year boy. In the sleep dynamics problem studied in this article, we take the impact of age on the sleep dynamics \cite{Vitiello2004} and EEG signal \cite{BoselliParrino1998,VanCauter2000} into account. Note that we only consider age since information provided in the available databases is limited.
In general, this approach can be understood as a high level filtering based on the {\em phenotype} information \cite{wu2018phenotype}.
Another benefit of this phenotype-based approach is its flexibility for the growing database. While it is widely believed that the larger the database is, the more accurate model we can establish, we should take the limited computational resource into account. By only choosing those subjects sharing similar phenotype to establish the prediction model, computational efficiency can be achieved.

This phenotype-based idea allows us to establish a ``self-evolving'' AI system for the automatic annotation purpose. Armed with the above ideas, the system can accumulate experience/wisdom from each new subject -- after applying the existing system with $n$ subjects with experts' annotations to the new subject to alleviate the physician's load, the physician can update/train the system by providing his/her feedback. The updated system with $n+1$ subjects is then more ``knowledgeable''. This close loop forms the self-evolving or self-learning part of the artificial intelligence system. See Figure \ref{fig:self-learning} for an illustration of the general framework.

To further validate and develop this proposed self-evolving system, in {our} future study we will include more {health-related} information, and mimic physicians' decision-making rules to determine the similarity between subjects. Moreover, we will establish a statistical model to better quantify the decision tree, and better handle the ``inter-physician'' variability of their decision trees. The ``reward'' idea from the reinforcement learning will also be considered.

\begin{figure}
\centering
\includegraphics[scale=0.45]{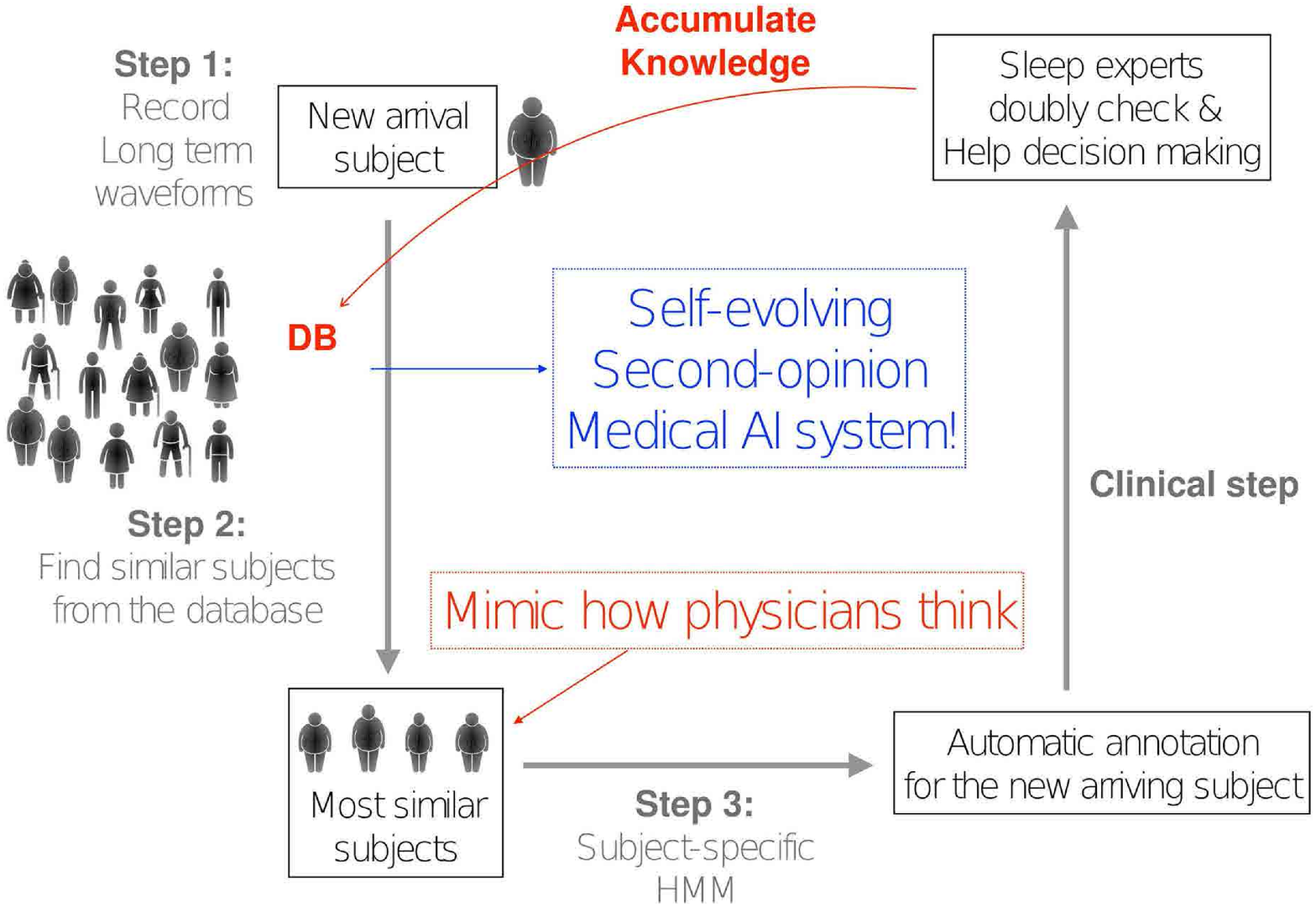}
\caption{An illustration of the ``Self-evolving'' artificial intelligence system.
}
\label{fig:self-learning}
\end{figure}

One main clinical application of this self-evolving AI system is establishing a more accurate sleep apnea screening. With the sleep stage information, the apnea-hypopnea index could be calculated. For that purpose, it is a common consensus to consider fewer channels for the patient to wear. While more channels provide more information, they may disturb the sleep. The respiratory flow is a common channel that people consider to screen the sleep apnea at home. In addition to {providing} the sleep apnea information, it has been known that the respiratory flow signal also contains abundant information about the sleep stage, {and} is based on different physiological {mechanisms} compared with the EEG signal \cite{Thompson2001}. In this scenario, the proposed sensor fusion algorithm has the potential to incorporate the information hidden in the flow signal and design a more accurate prediction system. An exploration of this direction will be postponed to our future work.

\subsection{Limitation and future work}
Despite the strength of the proposed method, the discussion is not complete without mentioning its limitations.
While we {test} the algorithm on the publicly available benchmark database, and compare our results with those of state-of-the-art algorithms, those databases are small. To draw a conclusion and confirm its clinical applicability, a large scale and prospective study is needed. We focus only on the spectral information in the EEG signals. There are other features, for example \cite{MOTAMEDIFAKHR201421}, we can consider to further improve the performance. While the spectral information is mainly determined by the nonlinear-type time-frequency representation SST for the purpose of preventing energy leakage, there are other time-frequency analysis tools that we can consider; for example, the scattering transform \cite{mallat2012group}. A systematic study of other possibilities will be explored in the future work.
While with two channels our algorithm {is compatible with} that reported in \cite{DeepSleepNet} {which} depends on only one channel, when we have only one channel, our algorithm does not perform better (For Fpz-Cz, the accuracy and F1 of our algorithm are 78.5\% and 67.9\%, while the accuracy and F1 reported in \cite{DeepSleepNet} are 82\% and 76.9\%. For Pz-Oz, the accuracy and F1 of our algorithm are 79.3\% and 70.3\%, while the accuracy and F1 reported in \cite{DeepSleepNet} are 79.8\% and 73.1\%). As discussed above, this limitation comes from the poor features for N1 classification. We need to find features that can better quantify N1 dynamics, and better understand how and why the deep neural network achieves the accuracy.
Another related open problem is how to further take the temporal information into account when we deal with the inter-individual prediction. Note that in the current algorithm, although the temporal relationship of epochs is considered in the HMM model, it is not taken into account to design the feature. The above mentioned limitations will be studied and reported in the future work.

Although extended theoretical understandings of applied algorithms have been established in the past decade, there are still open problems we need to explore from the theoretical perspective.
In general, we know that by taking the phase information into account, we obtain a sharper time-frequency representation. Although we do empirically find that the classification performance is better with the sharpened time-frequency representation determined by SST, we should be careful that a sharper time-frequency representation is not equivalent to the ``correct'' time-frequency representation. A mathematical question to ask is when there is no obvious oscillatory pattern in the EEG signal, like the ``mixed frequency'' property of the theta wave in N1, how does SST behave, and what is the mathematical property of the time-frequency representation determined by SST?
While AD and co-clustering look similar, they are developed under different motivations, and the consequence and relationship are never discussed. Understanding this relationship might allow us to further improve diffusion-based sensor fusion algorithms.

\section{Funding}
G.-R. Liu is supported by Ministry of Science and Technology (MOST) grant \texttt{MOST 106-2115-M-006 -016 -MY2}.
Y.-L. Lo is supported by MOST grant \texttt{MOST 101-2220-E-182A-001, 102-2220-E-182A-001, 103-2220-E-182A-001 and MOST 104-220-E-182-002}.
Y.-C. Sheu is supported by MOST grant \texttt{MOST 106-2115-M-009-006} and NCTS, Taiwan.

%\bibliographystyle{elsarticle-num}
%\bibliography{reference}
\end{document}